\documentclass[preprint]{aastex}

\shorttitle{Short-period planets} 

\shortauthors{Ida and Lin}

\begin{document}

\title{Towards a deterministic model of planetary formation. III.
Mass distribution of short-period planets around stars of various masses}

\author{S. Ida}
\affil{Tokyo Institute of Technology,
Ookayama, Meguro-ku, Tokyo 152-8551, Japan}
\affil{UCO/Lick Observatory, University of California, 
Santa Cruz, CA 95064}
\email{ida@geo.titech.ac.jp}

\and 

\author{D. N. C. Lin}
\affil{UCO/Lick Observatory, University of California, 
Santa Cruz, CA 95064}
\email{lin@ucolick.org}

\begin{abstract}
The origin of a recently discovered close-in Neptune-mass planet
around GJ436 poses a challenge to the current theories of planet
formation.  Based on the sequential accretion hypothesis and the
standard theory of gap formation and orbital migration, we show that
around M dwarf stars, close-in Neptune-mass ice-giant planets may be 
relatively common, while close-in Jupiter-mass 
gas-giant planets are relatively rare.  The mass distribution of 
close-in planets generally has two peaks at about Neptune mass 
and Jupiter mass.
The lower-mass peak takes the maximum frequency for M dwarfs.
Around more massive solar-type stars (G dwarfs), the higher-mass peak is 
much more pronounced.
These are because planets tend to undergo type II migration after
fully accreting gas around G dwarfs while they tend to migrate faster
than gas accretion around M stars.
Close-in Neptune-mass planets may also exist around G dwarfs, 
though they tend to be mostly
composed of silicates and iron cores and their frequency is expected to
be much smaller than that of Neptune-mass planets around M dwarfs and
that of gas giants around G dwarfs.  We also show that the conditions for
planets' migration due to their tidal interaction with the disk and
the stellar-mass dependence in the disk-mass distribution can be
calibrated by the mass distribution of short-period planets around
host stars with various masses.
\end{abstract}


\section{Introduction}
\label{sec:introduction}

In an attempt to place quantitative constraints on models of planet
formation, we developed an algorithm to simulate the kinematic
properties of gas giants formed in isolation (Ida \& Lin 2004a,
hereafter Paper I).  This prescription is based on the sequential
accretion model in which we assume that Jupiter-mass gas-giant planets
formed through 1) grain condensation, 2) runaway planetesimal coagulation
\citep{Greenberg78, WS89, ALP93, KI96}, 3)
oligarchic growth of protoplanetary embryos \citep{KI98,KI00}, and
4) gas accretion onto solid cores (embryos) 
\citep{Mizuno80, BP86, P96, Ikoma00}.  
Based on a distribution
of 1) dust-disk masses ($M_{\rm d}$) inferred from mm data \citep{BS96},
2) a range (1-10 Myr) of disk depletion time scale
($\tau_{\rm dep}$) inferred from the observed decline in the IR
\citep{HLL01} and mm data \citep{Wyatt}, 
and 3) three different growth-termination criteria \citep{LP93,
Bryden99}, we simulated a
distribution of protoplanetary masses.  In addition, we considered the
effect of post formation type-II orbital migration due to planet-disk
interaction \citep{LP85} which is an important process in
relocating protoplanets away from their birth places.  The last
process has been invoked \citep{Lin96} to account for the
origin of a population of Jupiter-mass short-period planets such as 51
Peg b \citep{MayorQueloz95}.

We compared the results of our simulation with the available data
of extrasolar planets.  We
suggested that around solar-type stars, 1) there may be a deficit of
intermediate-mass ($\sim 20-100 M_\oplus$) and intermediate-period
($\sim 0.1-1$yr)
in their mass-period distribution (Paper I), 
2) the frequency of gas giant planets may be an increasing function of their
host stars' metallicity [Fe/H]$_\ast$ (Ida \& Lin 2004b, hereafter
Paper II), and 3) a large fraction of the planets migrated to the proximity
of their host stars may have perished (Paper I, II). The first conclusion results
directly from the expectation that a) the growth of protoplanetary
cores is limited by dynamical isolation in the inner regions of
planetary systems and slow coagulation rate in the outer regions, 
b) even under the most favorable locations, $0.1-1$ Myr is needed for the formation of
protoplanetary embryos (cores), anywhere in the disk, with masses 
$M_{\rm c} \ga M_{\oplus}$, c) cores with mass larger than $M_{\rm c,acc} \sim$
several $M_\oplus$ can undergo runaway gas accretion,
and d) orbital migration occurs on a
similar time scale ($\tau_{\rm mig}$) to gas depletion in the disk
($\tau_{\rm dep}$).

The model we have presented so far provides the first step in the
construction of deterministic properties of planetary formation.  Its
simplistic assumptions must be re-examined in a wider context of
protoplanetary environment.  For example, planets are assumed to form
independently and in isolation in the model.  
Dynamical isolation in this context means that there are no other 
major planets around the same host stars. 
Most importantly,
post-formation dynamical interaction between the planets have not yet
been considered.  While these effects will be examined in a future
investigation, we consider here the mass function of 
dynamically-isolated planets and its dependence on 
the mass of the host star.  

\begin{figure}
\plotone{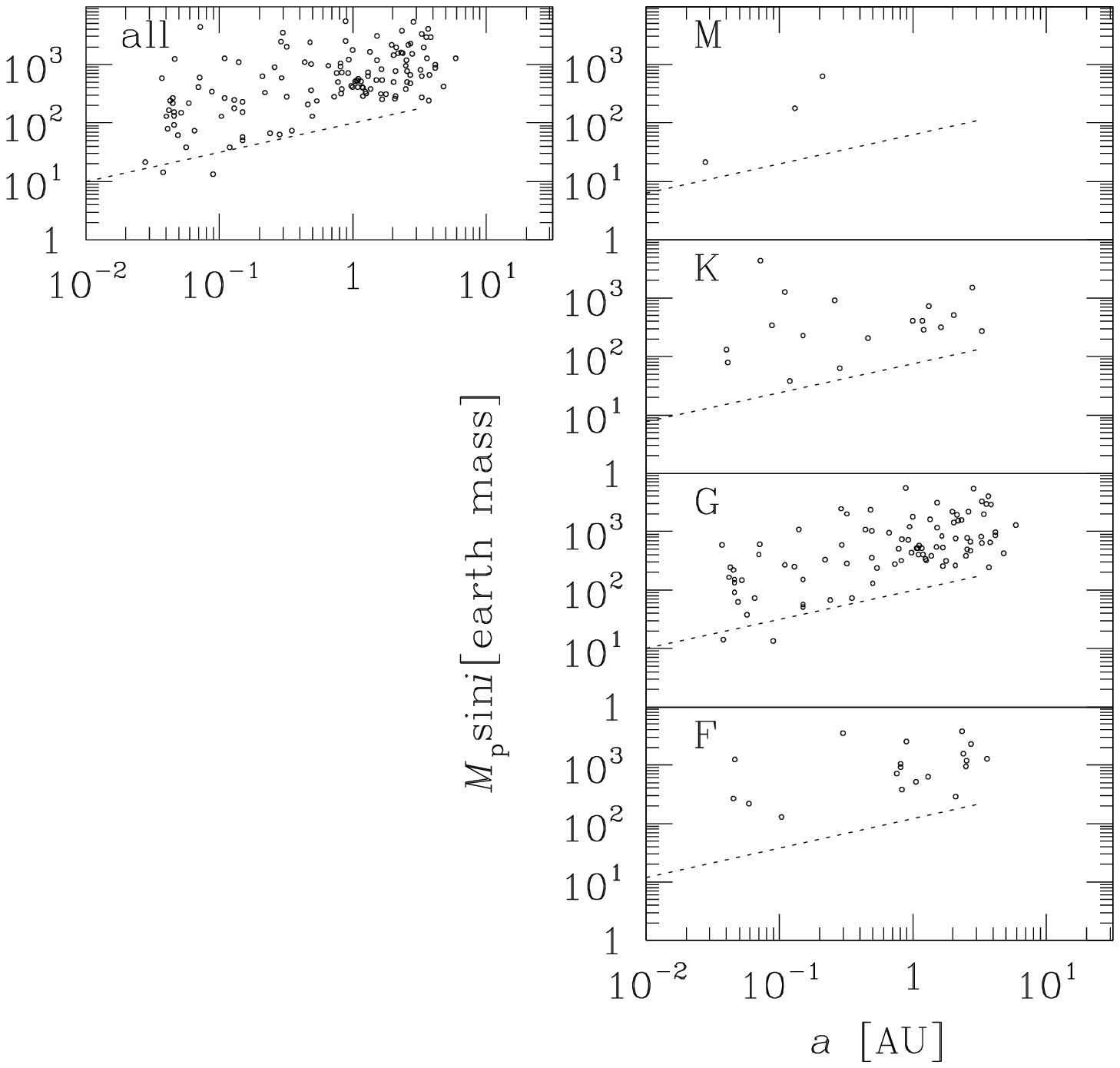}
\caption{
The distributions of semimajor axis ($a$) 
and mass ($M_{\rm p} \sin i$) of discovered extrasolar planets.
Unit of mass is Earth mass $M_{\oplus}$
(Jupiter mass is $M_{\rm J} \simeq 320M_{\oplus}$).
The data are taken from ``The Extrasolar Planets Encyclopedia''
(http://cfa-www.harvard.edu/planets/) as of February, 2005.
The planets around subgiants and those discovered by
transit survey are excluded.
The right panels show the planets
around M, K, G, and F stars, respectively.
The left panel shows all the data.
The dotted lines show observational limits ($v_r = 10$m/s)
for doppler survey. 
Detection of larger $a$ ($\ga 3$AU) planets, which have
longer orbital periods, is also limited.
}
\label{fig:obs}
\end{figure}

On the observational side, most of the known planets are found around
solar-type stars (G dwarf stars) because the most successful radial velocity 
surveys have been conducted for these target stars which have cool and well
defined atmospheric spectroscopic features. However, the search window
is rapidly expanding to lower-mass stars as detection techniques are
being refined in both radial velocity and transit searches.  
Figure \ref{fig:obs} shows the distributions of semimajor axis ($a$) 
and mass ($M_{\rm p} \sin i$) of discovered extrasolar planets
around M, K, G, and F stars.
The planets around subgiants are excluded, because
the relation between stellar spectral-type and its mass is
different from that for main sequence stars. 
The planets discovered by transit survey are also excluded,
because the transit survey has different observational bias 
in planetary periods from that of doppler survey.
Although much more planets have been discovered 
around G stars than around F and K stars,
the detection probability (after correction of metallicity
dependence) is similar among these stars (Fischer \& Valenti 2005).
However, the detection probability may be 
by order of magnitude lower around M stars.
Since their signals are the most conspicuous, close-in planets are expected
to be the first to be uncovered (Narayan {\it et al.} 2004).
However, until recently, the only planets discovered
around M stars are two Jupiter mass planets with moderate
semimajor axis (0.13AU, 0.21AU) around Gliese 876.  
No Jupiter-mass close-in planets have been discovered around M stars.

Recently, a short-period (2.6 days) Neptune-mass ($21 M_\oplus$)
planet is found to be orbiting around an M-dwarf ($M_\ast = 0.4
M_{\odot}$) GJ436 \citep{Butler04}.
With a mass between those of gas giants and the Earth, this finding
signifies a transition in the quest to search for terrestrial planets.
In the solar system, two ice giants, Uranus and Neptune, have masses
in this range.  These planets are primarily composed of icy cores with
a modest gaseous envelope.  In accordance with the core accretion
scenario, e.g., \citep{Wuchterl00}, 
the cores in the outer solar system take a long time to
emerge and when they finally acquired $M_{\rm c,acc}$, the solar nebula
was already so severely depleted that they can only accreted a small
amount of gaseous envelope \citep{Hayashi85}.  It is natural
to extrapolate that GJ436b may also have attained $M_{\rm c,acc}$ but
failed to accrete much gas.  The main challenge to such a scenario is
to account for the origin of both its low mass and short period.

The period of the first extrasolar gas giant planet (51 Peg b) discovered 
\citep{MayorQueloz95} around a main sequence star outside the solar system
is comparable to that of GJ436b.  That planet and dozens others like
it are thought to have formed through sequential accretion beyond the
ice boundary and migrated to their present locations \citep{Lin96} 
as a consequence of their tidal interaction with their nascent
gaseous disks \citep{GT80, LP85}.
But, the much lower mass $M_{\rm p}$ of GJ436b implies that gas accretion
onto it may have been greatly suppressed prior to, during, and after
its migration.

Two other short-period Neptune-mass 
planets have been found around G dwarfs (a planet with
2.8 day period and $15M_\oplus$ mass around 55 Cnc 
and that with 9.6 day period and $14M_\oplus$ mass around HD 160691)
\citep{McArthur04, Santos04}.  
Since 55Cnc and HD160691 have three and two other giant 
planets (maybe gas giants), these Neptune-mass planets could 
form {\it in situ} by accumulation of rocky materials
caused by sweeping mean motion resonance associated with
migration of a giant planet, e.g., \citep{Malhotra, Ida00} 
or sweeping secular
resonance associated with disk gas depletion,
e.g., \citep{Ward81, Nagasawa00, Lin_etal04}. 
Since we do not include interaction from other major planets, 
formation of these Neptune-mass planets is beyond scope of
the present paper.
On the other hand, no gas giant planets has been
found around GJ436.
GJ436b is dynamically-isolated, so that
its formation must be considered without help of
other giant planet(s) and our calculation in the present paper
can address its formation.

Although in the present paper, we will propose a
scenario that type II migration of a Neptune-mass planet occurs 
without significant gas accretion onto the planet around an M star,
there are several other potential scenarios for the origins of
dynamically-isolated, close-in Neptune-mass planets. 
\citet{Boss_etal02} suggested
that, along with Jupiter and Saturn, Uranus and Neptune were formed through
gravitational instability, in a massive disk.  
Heavy elements settled to form the cores and the gas envelope (clump) 
was greatly depleted by the photo-evaporation due to the UV flux from 
nearby OB stars before it contracts to planetary size.  
Since gravitational instability is unlikely
at the present location (0.028AU) of GJ436b, the planet must have
migrated from outer region by tidal interaction of the disk.
However, the gravitational
potential of disks around M stars is shallower than that around 
G stars such that the evaporation of the envelope of collapsing clumps
would also eliminate all the residual gas in the disk.  
It would be difficult for the collapsing fragment to lose a large fraction of 
their mass and migrate extensively to form a GJ436b-like planet.
On the other hand, under some extreme circumstance, the sporadic
UV and X-ray irradiation from its host star 
could evaporate envelope of a gas giant planet that has migrated 
from outer region during M star's main sequence lifetime.
But the overall impact of the photoevaporation process on 
planet's envelope and mass has not been determined.

Under the general concept of sequential accretion scenario, 
dynamically-isolated close-in
planets may also form {\it in situ} \citep{Bodenheimer_etal00}.  This
scenario requires a concentration of planetesimals in the stellar
proximity.  One possible mechanism which may lead to such a situation
is through embryo-disk interaction, or commonly known as type-I
migration \citep{Ward86, Ward97}.  
The accumulation
of building blocks for close-in planets also requires the termination of
their migration process and the interaction between the embryos with 
their host stars as well as residual planetesimals.  

Through a case study, we consider in this paper the origin of 
dynamically-isolated close-in planets of
$\sim$ Neptune mass around stars with various masses.  
With the model developed in Paper I and II, we show that
Neptune-mass close-in planets may be abundant around M stars.
In \S2, we briefly recapitulate the sequential accretion hypothesis.  Using
our prescription, we first compare the mass function of
dynamically-isolated close-in versus modest to long-period planets
around solar-type stars (G dwarfs) in \S3.  
This comparison is useful because we
have most observational data and constraints for planets around 
solar-type stars at the present moment.  Based on the best available
observed properties of pre-main sequence evolutionary tracks of
different-mass stars and accretion rates onto their host
stars, we construct disk models around main sequence stars with
various masses.  In \S4, we construct a conventional model of giant
planet formation around lowest-mass stars and show that 
icy planets with 10-$20 M_{\oplus}$ accrete from planetesimals at $\sim 1$AU 
without any significant gas accretion onto the planets.  
These Neptune-mass planets can also readily migrate to
the proximity of the stellar surface, where gas accretion is
quenched at Neptune-mass.
Through a series of
simulations, we show, in \S5, that the disk mass dependence
on their host stars' mass and the criteria of tidally induced
migration in the disk may be quantitatively constrained by the mass
distribution of short-period planets.  
We also place constraints, in \S6, on the
dependence of disk mass on the stellar mass.  Finally, in \S7, we
summarize our results and discuss their implications.

\section{Core accretion in disks around low-mass stars}
\label{sec:core}

The detailed description of sequential accretion scenario and our
prescription to simulate the formation of planets are given in Paper I.
We briefly recapitulate the central features of our approach and
define various quantities which are used in the discussions of our
results.
 
\subsection{Growth of protoplanetary embryos}

In the sequential accretion scenario, planetesimals grow into
protoplanetary embryos (cores) which affects the velocity dispersion
$\sigma$ of nearby planetesimals and modifies their own growth
\citep{IM93, ALP93, Rafikov03}.
But, $\sigma$ is also affected by gas drag \citep{KI02}.  In a
disk with a surface density of dust ($\Sigma_{\rm d}$) and gas
($\Sigma_{\rm g}$) around a host star with mass $M_\ast$, 
protoplanetary embryos' mass at any location $a$
and time $t$ is
\begin{equation}
M_{\rm c}(t) \simeq 
\left( \frac{t}{0.48 {\rm Myr}} \right)^3 
\left( \frac{\Sigma_{\rm d}}{10 \mbox{gcm}^{-2}} \right)^{3}
\left( \frac{\Sigma_{\rm g}}{2.4 \times 10^3 \mbox{gcm}^{-2}} \right)^{6/5}
\left( \frac{m}{10^{22}{\rm g}} \right)^{-2/5}
\left( \frac{a}{1{\rm AU}} \right)^{-9/5} 
\left( \frac{M_\ast}{M_{\odot}} \right)^{1/2} M_{\oplus},
\label{eq:m_grow}
\end{equation}
where $m$ is the typical mass of the planetesimals accreted by the embryos.  
Since accretion time $\tau_{\rm c,grow} = M_{\rm c}/\dot{M}_{\rm c}$ 
increases with $M_{\rm c}$, $M_{\rm c}(t)$ does not depend on its initial
value $M_{\rm c,0}$ as long as $M_{\rm c}(t) \gg M_{\rm c,0}$. 

In the limit of small $\sigma$, the full width of embryos' feeding
zone ($\Delta a_{\rm c}$) is limited to $\sim 10 r_{\rm H}$ 
\citep{Lissauer87, KI98} where $r_{\rm H}$ is the
embryos' Hill's radius $r_{\rm H}$ $(= (M_{\rm c}/3M_\ast)^{1/3}a)$.  
When all
the residual planetesimals in an embryo's feeding zone have coagulated
with it, the embryo attains an isolation mass
\begin{equation}
M_{\rm c,iso} \simeq
0.16 \left(\frac{\Sigma_{\rm d}}{10 \mbox{gcm}^{-2}}\right)^{3/2}
\left(\frac{a}{1\mbox{AU}}\right)^{3}
\left(\frac{\Delta a_{\rm c}}{10r_{\rm H}} \right)^{3/2} 
\left( \frac{M_\ast}{M_{\odot}} \right)^{-1/2} M_{\oplus}.
\label{eq:m_iso}
\end{equation}

\subsection{Protostellar disk properties}

Equations (\ref{eq:m_grow}) and (\ref{eq:m_iso}) indicate that both
$M_{\rm c}(t)$ and $M_{\rm c, iso}$ are determined by the distribution of
$\Sigma_{\rm d}$ and $\Sigma_{\rm g}$.  In Paper I and II, we introduced a
multiplicative factor ($f_{\rm d}$ and $f_{\rm g}$) to globally scale 
disks with the minimum mass
nebula model for the solar system \citep{Hayashi81} such that
\begin{equation}
\left\{ \begin{array}{ll} \Sigma_{\rm d} & = 10 \eta_{\rm ice} 
f_{\rm d} h_{\rm d} (a/ {\rm 1 AU})^{-3/2} \;\; [{\rm g~cm}^{-2}], \\ \Sigma_{\rm g} 
\label{eq:sigma_dust}
&
= 2.4 \times 10^3 f_{\rm g} h_{\rm g} (a/ {\rm 1 AU})^{-3/2} \;\; [{\rm g~cm}^{-2}],
\label{eq:sigma_gas}
\end{array} \right.
\end{equation}
where the step function $\eta_{\rm ice} = 1$ inside the ice boundary at
$a_{\rm ice}$ and 4.2 for $a \ge a_{\rm ice}$.
[Note that the latter can be slightly
smaller ($\sim 3.0$) \citep{Pollack94}.]
The minimum mass model corresponds to $f_{\rm d} = f_{\rm g} \sim 1$.
Here we introduced a new scaling factor, $h_{\rm d}$ and $h_{\rm g}$, 
representing the dependence on stellar mass (see below).

If the disk is optically thin and heated
by stellar irradiation only \citep{Hayashi81},
\begin{equation}
a_{\rm ice} = 2.7 (L_\ast/L_\odot)^{1/2} {\rm AU}.
\end{equation}
The stars' luminosity $L_\ast$ is generally a function of their mass
$M_\ast$ and age $t$.  
Additional heating due to viscous
dissipation enlarges this boundary \citep{LP80}.  This
{\it ad hoc} phenomenological prescription provides a useful working
hypothesis for comparative analysis between solar system architecture
and extrasolar planetary systems.
Since $L_{\ast}$ is generally an increasing function of $M_{\ast}$,
$a_{\rm ice}$ is small in disks around low-mass stars. 

Around solar-type T Tauri stars, the observationally inferred total
mass of dust, $M_{\rm d}$, in the protostellar disks ranges from
$10^{-5} M_{\odot}$ to $10^{-3}M_{\odot}$ \citep{BS96}, which corresponds to a
range of $f_{\rm d} \sim 0.1-10$.  The disk-mass
determinations are somewhat uncertain due to the poorly known
radiative properties of the grains.  Nevertheless, their divergent
dust content provides a reasonable evidence for a greater than an
orders of magnitude dispersion in $M_{\rm d}$.  In addition, 
the radio image
of the disks is not well resolved in many cases.  A rough magnitude of
$f_{\rm d}$ can be inferred from the the total mass of the disk under
the assumption that all disks have similar sizes (a few tens AU to a
hundred AU).  In this paper we follow Paper II and generate a set of
$f_{\rm d}$ with a unit variant of Gaussian logarithmic distributions
and a range between 0.1-10 
with cut-off of high $f_{\rm d} h_{\rm d}$ tails at $30 (M_\ast/M_\odot)$
(see Figure \ref{fig:f_d_dist}).
If $f_{\rm g}h_{\rm g} \ga 30 (M_\ast/M_\odot)$, the disk is gravitationally
unstable at $a \ga$ a few AU (see discussion in section 7), 
and the total mass of heavy
elements in the disk would be significant fraction of that in the
host star.

In eq.~(\ref{eq:sigma_dust}),
the dependence of $\Sigma_{\rm d}$ on $M_\ast$ is incorporated in a
mass scaling function $h_{\rm d}$.  In general, a relatively weak IR
excess is associated with low-mass stars, which suggests $h_{\rm d}$
may be an increasing function of $M_\ast$, at least in the inner
regions of the disk.  For host stars with $M_\ast = M_\odot$,
$h_{\rm d} = 1$.

Equation (\ref{eq:m_grow}) indicates that $M_{\rm c}$ is also a
function of $\Sigma_{\rm g}$ through damping of $\sigma$ due to
gas drag.  
The planetary migration rate that we mention later is also dependent
on $\Sigma_{\rm g}$.  
Around other stars, very little information is available on
$\Sigma_{\rm g}$.  
Since there is no indication of divergent depletion
pattern between molecular hydrogen and mm-size dust emission
\citep{Thi01}, 
we follow the same prescription for $\Sigma_{\rm d}$
with a disk mass scaling parameter $f_{\rm g}$ and a stellar-mass
dependence function $h_{\rm g}$ (eq.~[\ref{eq:sigma_dust}]).
Similar to Paper I and II, we adopt the conjecture that $f_{\rm d}$ does not 
change except in those regions where they have been totally accreted by the
cores and 
\begin{equation}
f_{\rm g} = f_{\rm g,0} {\rm exp} \left( - {t \over \tau_{\rm dep}} \right),
\label{eq:fg_tdep}
\end{equation}
where $\tau_{\rm dep}$ is disk depletion time scale discussed below.
The assumption of the uniform exponential decay is for simplicity.
The effects of detailed decay pattern as a result of viscous
evolution of disks will be discussed in a separate paper. 
For computational simplicity, we
assume [Fe/H]=0, {\it i.e.}, a solar composition for all stars so that
$f_{\rm d} = f_{\rm g,0}$ and $h_{\rm g} = h_{\rm d}$. The dependence of
planetary $M_{\rm p}-a$ distribution on [Fe/H] for solar-type stars
has already been discussed in Paper II.  

We now consider a prescription for the stellar mass dependence.
The parameters are $h_{\rm d}$ ($= h_{\rm g}$) and $\tau_{\rm dep}$.
In young clusters, the
fraction of stars with detectable IR \citep{HLL01} and mm continuum
\citep{Wyatt} from circumstellar disks around T Tauri stars
declines on the time scales of 1-10 Myr.  Although this decline may be
due to dust growth and planetesimal formation rather than the
depletion of heavy elements \citep{DAlessio01, Tanaka05},
the correlation between the intensity of mm dust continuum with
the gas decline emission \citep{Thi01} and the UV veiling for ongoing 
gas accretion suggest that gas is depleted as the dust signature fades.
The dust signature maintains up to 10Myr also for disks around 
brown dwarfs \citep{Mohanty03}.
So, in this paper, we assume $\tau_{\rm dep}$ in the range of 1-10 Myr
for all stellar masses.

Although direct estimates of disk mass both in gas and dust are difficult 
to obtain for the inadequate sensitivity of existing observational
instruments, the disk accretion rate $\dot M$ can be inferred from 
the H$\alpha$ line profiles \citep{Muzerolle03, Natta04} such that 
$\dot M \propto M_\ast ^2$ with a large dispersion.  
If the angular momentum
transfer and mass diffusion time scale is insensitive to $M_\ast$, we
could infer $h_{\rm g}$ $(= h_{\rm d})$ $\sim (M_\ast/M_\odot)^2$.  
In view of the large
uncertainty in the data, we consider
three possible dependence on $M_\ast$,  
\begin{equation}
h_{\rm d} = (M_\ast/M_\odot)^{0, 1, 2},
\label{eq:hd}
\end{equation}
with $h_{\rm d} = (M_\ast/M_\odot)^2$ as a standard case.

\subsection{Core growth, isolation mass, and gas giant formation 
around stars with different masses} 

With these prescriptions for disk parameters, we find that
\begin{equation}
M_{\rm c}(t) \simeq
\left( \frac{t}{4.8 \times 10^5{\rm years}} \right)^3 \eta_{\rm ice}^{3}
f_{\rm d}^{3} h_{\rm d}^{3} f_{\rm g,0}^{6/5} h_{\rm g}^{6/5}
\left( \frac{a}{1{\rm AU}} \right)^{-81/10}
\left(\frac{M_\ast}{M_{\odot}} \right)^{1/2}
M_{\oplus},
\label{eq:m_grow0}
\end{equation}
by assuming $m=10^{22}$g.  
From eq.~(\ref{eq:sigma_dust}), we estimate
the core masses in protoplanetary systems with eq.~(\ref{eq:m_iso})
such that
\begin{equation}
M_{\rm c,iso} \simeq
0.16 \eta_{\rm ice}^{3/2} f_{\rm d}^{3/2} h_{\rm d}^{3/2}
\left(\frac{a}{1\mbox{AU}}\right)^{3/4} 
\left(\frac{M_\ast}{M_{\odot}} \right)^{-1/2} M_{\oplus}.
\label{eq:m_iso0}
\end{equation}
We have already pointed out in Paper I and II that the growth of
planetesimals is limited by isolation at small $a$ and the slow
coagulation rate at large $a$.  
Since for smaller $M_{\ast}$, $r_{\rm H}$ is larger, 
$M_{\rm c,iso}$ is larger for the same mass
disks.  However, both $h_{\rm d}$ and $h_{\rm g}$
are increasing functions of $M_\ast$, so that the growth of $M_{\rm c} (t)$
and the isolation mass $M_{\rm c, iso}$ at any given $a$ are actually smaller 
for lower-mass stars.
We do not explicitly include type I migration \citep{Ward86, Ward97} 
of cores, but take into account its effects in some runs 
(see Paper I and discussion in \S5).

As the cores grow beyond a mass 
\begin{equation}
M_{\rm c,hydro} \simeq
10 \left( \frac{\dot{M}_{\rm c}}{10^{-6}M_{\oplus}/ {\rm yr}}\right)^{0.25}
M_{\oplus},
\label{eq:crit_core_mass}
\end{equation}
their planetary atmosphere is no longer in hydrodynamic equilibrium
and they begin to accrete gas \citep{Stevenson82, Ikoma00}.  
In the above equation, we neglected the dependence on opacity (see Paper I).
In regions where they have already attained isolation, the cores'
accretion $\dot M_{\rm c}$ is much diminished and 
$M_{\rm c,hydro}$ can be comparable to an Earth mass.  But the gas accretion rate
is still regulated by the efficiency of radiative transfer such that
\begin{equation}
\frac{dM_{\rm p,g}}{dt} \simeq \frac{M_{\rm p}}{\tau_{\rm KH} }
\label{eq:mgsdot}
\end{equation}
where $M_{\rm p}$ is the planet mass including gas envelope
and the Kelvin Helmholtz contraction time scale is 
(for details, see Paper II)
\begin{equation}
\tau_{\rm KH} \simeq 10^{10} \left(\frac{M_{\rm p}}{M_{\oplus}}\right)^{-3} \; {\rm yrs}.
\label{eq:tau_KH}
\end{equation}

Gas accretion onto the core is quenched when the disk is depleted
either locally or globally.  A protoplanet induces the opening of a
gap when its rate of tidally induced angular momentum exchange with
the disk exceeds that of the disk's intrinsic viscous transport \citep{LP85}, 
that is, when the planet mass $M_{\rm p}$ exceeds
\begin{equation}
M_{\rm g,vis} \simeq \frac{40 \nu}{a \Omega_{\rm K}} M_\ast 
\simeq 40 \alpha \left(\frac{h}{a}\right)^2 M_\ast
\simeq 3 \left(\frac{\alpha}{10^{-4}}\right)
\left(\frac{a}{1{\rm AU}}\right)^{1/2}
\left(\frac{L_\ast}{L_{\odot}}\right)^{1/4} M_{\oplus},
\label{eq:m_gas_vis}
\end{equation}
where we used an equilibrium temperature in optically thin disks
\citep{Hayashi81} and $\alpha$-prescription for
the effective viscosity $\nu$ (Shakura \& Sunyaev 1973), in which $T=
170 (a/a_{\rm ice})^{-1/2}$ K and $\nu = \alpha h^2 \Omega_{\rm K}$
where $\alpha$ is a dimensionless parameter, $h$ and
$\Omega_{\rm K}$ are the disk scale height and Kepler frequency. 
Since $h = c_{\rm s}/\Omega_{\rm K}$ where $c_{\rm s}$ is the sound
velocity, $h^2 \propto T/M_{\ast} \propto L_{\ast}^{1/4}/M_{\ast}$,
so that $M_{\rm g,vis} \propto L_{\ast}^{1/4}$.
Since $L_{\ast}$ is an increasing function of $M_{\ast}$,
eq.~(\ref{eq:m_gas_vis}) indicates that $M_{\rm g,vis}$ 
is smaller around lower-mass stars. 
The orbital
evolution of planets is locked to the viscous evolution of the disk
gas (type II migration) 
when their $M_{\rm p} \ge M_{\rm p,mig} = A_\nu M_{\rm g, vis}$ 
where $A_\nu$ is the dimensionless factor
$\sim 3-10$ for a laminar disk \citep{LP85}.

The type II migration rate $\dot a$ is given by $a/ \tau_{\rm mig}$ with
\begin{equation}
\tau_{\rm mig} = 10^6 f_{\rm g}^{-1} h_{\rm g}^{-1}
\left(\frac{\alpha}{10^{-4}}\right)
\left(\frac{M_{\rm p}}{M_{\rm J}} \right)
\left(\frac{a}{1{\rm AU}}\right)^{1/2} \;{\rm yrs},
\label{eq:tau_mig}
\end{equation}
where $M_{\rm J}$ is Jupiter mass.
We set a lower limit on $\tau_{\rm mig}$ at 
$a^2/\nu \sim 4.3 \times 10^5 (\alpha/10^{-4})(a/1{\rm AU})$ yrs (Paper I).
We found that the formula for $\dot{J}_m$ 
(angular momentum flux at the radius of maximum viscous couple, $r_m$) 
given in Paper I (eq.~[63] in the paper) must be multiplied 
by a factor $2 \pi$.
If we use $\dot{J}_m$ to evaluate evolution of planetary orbital 
radius $a$, eq.~(\ref{eq:tau_mig}) is reduced by $2 \pi$ as well
(Paper I).  However, since the planetary migration may be
caused by a fraction of $\dot{J}_m$ and the fraction is 
uncertain, we use eq.~(\ref{eq:tau_mig}) in the present
paper too.  We adjust the migration time scale by the value of $\alpha$, 
which is also uncertain, comparing with observational data. 
As shown in Paper II, in order to reproduce period distribution of
gas giant planets similar to that of observed 
extrasolar planets around solar-type stars, 
the disk viscous diffusion time scale $\tau_\nu = r_m^2/\nu
\sim 4 \times 10^6 (\alpha/10^{-4})(r_m/10{\rm AU})$ yrs 
must be comparable to disk lifetime $\tau_{\rm dep}$.
Hence, the $\alpha$ viscosity in our model must be $\sim 10^{-4}$.
Although the regions at 1-10AU could be "dead zone" for
MHD turbulence \citep{Sano00}, resulting in a very small $\alpha$ 
in these regions, the best fit value 
$\alpha \sim 10^{-4}$ may not necessarily reflect a realistic value
because of rather simple assumptions for $\Sigma_g$ distribution 
and its exponential decay in our model.
We will carry out more detailed calculations coupled with
disk viscous evolution in a separate paper.

The planets' migration is terminated either when the disk is severely
depleted ($f_{\rm g} \rightarrow 0$) or when they reach $a_{\rm stall}$,
which is set to be 0.04AU in our calculations.
There are potential mechanisms to stop migration at $\la 0.05$AU
\citep{Lin96}.  However, Paper II suggests that only a small 
faction ($\la 10$\%)
of migrating planets can survive in the vicinity of their host stars.
This fact should be kept in mind when population of close-in
planets is discussed with our model.

The gap becomes locally severely depleted when the planets' Hill's
radius ($r_{\rm H}$) exceeds the disk thickness $h$ 
\citep{Bryden99}, that is, when
$M_{\rm p}$ exceeds
\begin{equation}
M_{\rm g,th}
\simeq 120 \left(\frac{a}{1{\rm AU}}\right)^{3/4}
\left(\frac{L_\ast}{L_\odot}\right)^{3/8}
\left(\frac{M_\ast}{M_\odot}\right)^{-1/2} M_\oplus.
\label{eq:m_gas_th}
\end{equation}
Growth through gas accretion is quenched for planets with 
$M_{\rm p} \ge M_{\rm p,trunc} = A_{\rm th} M_{\rm g,th}$.
Numerical simulations show some uncertainties in
the dimensionless parameter $A_{\rm th}$ \citep{Bryden99, Nelson00}.  Planets with 
$M_{\rm p,trunc} > M_{\rm p} > M_{\rm p,mig}$ migrate 
with the disk while continue to accrete gas, albeit at a reduced rate,
e.g., \citep{Lubow99}. 
We use eq.~(\ref{eq:mgsdot}) without a reduction factor for simplicity,
because the reduction factor 
is quite uncertain and introduction of the factor does not affect
the results significantly.
Since gas accretion for $M_{\rm p} > M_{\rm p,mig}$ is
already very rapid (\ref{eq:tau_KH}), 
the reduction does not change total gas accretion time scale.
Following Paper I, we adopt
in this paper $A_{\rm th} =1.5^3 \simeq 3.4$, that is, the truncation condition
is $r_{\rm H} > 1.5h$.
Since $L_{\ast}$ rapidly increases with $M_{\ast}$,
$M_{\rm p,trunc}$ is also smaller around lower-mass stars. 

Even for planets with $M_{\rm p} < M_{\rm p,trunc}$, gas accretion may be
ultimately limited by the diminishing amount of residual gas in the
entire disk.  For our disk models, the maximum available mass is
\begin{equation}
M_{\rm g,no iso} \sim \pi a^2 \Sigma_{\rm g} 
\simeq  290 f_{\rm g,0} h_{\rm g} 
\left(\frac{a}{1\mbox{AU}}\right)^{1/2} {\rm exp} 
\left(-\frac{t}{\tau_{\rm dep}}\right) M_{\oplus}.
\label{eq:m_gas_non_iso}
\end{equation}
When $M_{\rm g,noiso}$ becomes smaller than $M_{\rm p}$, gas accretion is
terminated.

A similar global limit $M_{\rm c, no iso} = \pi a^2 \Sigma_{\rm d}$ 
(see eq.~[\ref{eq:m_noiso_sp}]) is 
also imposed if the formula (\ref{eq:m_iso}) exceeds it. 
As the gas is severely depleted, the velocity dispersion $\sigma$ of 
the embryos and residual
planetesimals grows until they cross each other's orbits
\citep{Iwasaki02, Kominami02}.  Eventually a few surviving embryos
acquire most of the residual planetesimals and less massive cores
during the late oligarchic-growth stage.  The asymptotic embryos'
masses are given by eq.~(\ref{eq:m_iso}) with $\Delta a_{\rm c} \sim
V_{\rm surf}/\Omega_{\rm K}$,
\begin{equation}
M_{\rm e,iso} \simeq
0.52 \eta_{\rm ice}^{3/2} f_{\rm d}^{3/2} h_{\rm d}^{3/2}
\left(\frac{a}{1{\rm AU}}\right)^{3/2}
\left(\frac{\rho_{\rm d}}{1{\rm gcm}^{-3}}\right)^{1/4} 
\left(\frac{M_\ast}{M_\odot}\right)^{-3/4} M_{\oplus}
\label{eq:m_e_iso}
\end{equation}
where $V_{\rm surf}$ and $\rho_{\rm d}$ are the surface escape speed 
and internal density of the embryo.  We use this enlarge asymptotic mass
when $f_{\rm g} < 10^{-3}$.

In Paper I and II, we put all of these processes into a numerical scheme
to simulate the formation and migration probabilities of planets around solar-type
stars. Cores with $M_{\rm p} > M_{\rm c,acc}$ emerge on time scale shorter
than $\tau_{\rm dep}$ in disks with modest-to-large values of 
$f_{\rm d}$ around solar-type stars. 
In this limit, gas accretion and orbital migration lead to the
formation of gas giants with kinematic properties similar to those
observed.

\section{Emergence and migration of Neptune-mass planets in disks around 
solar-type stars}
\label{sec:icegiant}

We now apply our numerical methods to study the formation of planets
around stars with various masses.  Our objective is to simulate the
mass function of close-in planets and assess the influence of
formation and migration on it. We show here that this quantity can
provide clues on the dominant processes which regulate planet
formation and it also can be used to distinguish between competing
theories of planet formation.  In all models, we choose $\alpha =
10^{-4}$ based on assumption that the viscous evolution time scale for
the disks is comparable to $\tau_{\rm dep}$ (see \S2.3).
Using these models, we carry out Monte Carlo simulations.
For simplicity, we generate a set of initial $a$'s of the
protoplanets and $\tau_{\rm dep}$ with uniform
distributions in log scale in the ranges of 0.1--$100$AU and 
$10^6$--$10^7$ yrs.
The assumed $a$ distribution corresponds to orbital
separations $\Delta a$ that is proportional to $a$, which may be
the simplest choice.
The distribution of $f_{\rm d}$ was discussed in \S2.2
(also see Figure \ref{fig:f_d_dist}).
In Paper I and II, we also assumed a distribution of $M_\ast$ in a range
of 0.7-1.4$M_\odot$.
In this paper, in order to make the $M_\ast$ dependence clear, 
the value of $M_\ast$ is fixed in each run.

In the following sections, we consider several sets of model parameters.
We first discuss a standard model with $A_{\nu} = 10$ and 
$h_{\rm d} = (M_\ast/M_\odot)^2$
around a solar-type star
with $M_{\ast} = 1.0 M_{\odot}$ (model 1.0).
The evolution of totally 20,000 planets are calculated for each run.

\subsection{Formation of cores and asymptotic mass of protoplanets.}

Model 1.0 is similar to the results we have already presented in Paper I.
In Figure \ref{fig:ma}a, we highlight the mass $(M_{\rm p,fin})$ 
and semi major axis $(a_{\rm fin})$ distribution of planets
at $t=10^9$ yrs after they have attained 
their asymptotic mass and gone through the
initial migration due to their tidal interaction with their nascent disks.
The main features to notice in this panel are: 1) a deficit of planets
with intermediate masses ($20-100 M_\oplus$) at intermediate
semi major axis ($0.1-1 {\rm AU}$), and 2)
a large population of close-in ($a_{\rm fin} < 0.05$AU) 
gas giants with $20 M_\oplus \la M_{\rm p}
\la 2 \times 10^3 M_\oplus$,
although only a small faction of them ($\la 10$\%) 
may be able to survive (Paper II).
We have already indicated in Paper I that
these properties are due to 1) the runaway nature of dynamical gas
accretion and 2) type II migration.

In order to distinguish between these two dominant effects, we trace
back, in Figure \ref{fig:ma}b, the initial semi major axis
$(a_{\rm ini})$ where the cores of both close-in (marked by black circles)
and intermediate or long-period planets (marked by gray dots) formed.
These results clearly indicate an one-to-one mapping between the mass
function of the close-in planets and the locations where they are
formed.  For illustrative purposes, we also mark the domain where some
physical processes operate and dominate the evolution of protoplanets.
For example, the thin solid lines indicate the upper limit of the
isolation mass that cores can attain prior to gas
depletion (it is obtained with $f_{\rm d} =30$ in $M_{\rm c, iso}$ 
and $M_{\rm c, no inso}$).
The transition at 2.7 AU in model 1.0 corresponds to the ice boundary. 
The thick solid lines
indicate the critical mass for the onset of type II migration, 
$M_{\rm p,mig} = A_\nu M_{\rm g, vis}$ with $A_\nu = 10$
according to eq.~(\ref{eq:m_gas_vis}).  
We also highlighted the asymptotic growth limit 
$M_{\rm p,trunc} = A_{\rm th} M_{\rm g,th}$ with $A_{\rm th} = 3.4$ 
(eq.~\ref{eq:m_gas_th}) by broken lines.

At any given $a_{\rm ini}$, a fraction of terrestrial planets can form
with $M_{\rm p}$ above $M_{\rm p,iso}$ through 1) gas accretion and 2)
merger of residual planetesimals and other embryos after disk gas depletion. 
When the embryos reach their
isolation mass, $M_{\rm c, hydro}$ declines with the vanishing 
$\dot M_{\rm c}$ (eq.~[\ref{eq:crit_core_mass}]).  
Although gas accretion is initiated, 
planets cannot grow significantly prior to severe gas depletion 
unless the
planets' isolation mass $M_{\rm c, iso} \ga M_{\rm c,acc} \simeq$ 
several $M_\oplus$.
The gas accretion time scale given by eq.~(\ref{eq:tau_KH})
can be comparable to or shorter than $\tau_{\rm dep}$ only 
for cores with $M_{\rm c} \ga$ several $M_\oplus$.
Mergers of residual planetesimals and other embryos
occur during and after the gas depletion (see eq.~[\ref{eq:m_e_iso}] for
asymptotic $M_{\rm c} \sim M_{\rm e,iso}$) (Kominami \& Ida 2002).  

If their $M_{\rm p} > M_{\rm p,mig} = A_\nu M_{\rm g, vis}$ and there is 
adequate residual gas in the disk ($f_g$ is still large enough), 
they would migrate to the vicinity of their
host stars (eq.~[\ref{eq:tau_mig}] with eq.~[\ref{eq:fg_tdep}]).
The magnitude of $M_{\rm g, vis}$ is an increasing
function of $a$.
For $a \la 0.7$ AU, $\tau_{\rm KH}$ for embryos
which can migrate is longer than their $\tau_{\rm mig}$.  Although
their $M_{\rm p}$ may be $< M_{\rm p,trunc} = A_{\rm th} M_{\rm g,th}$, 
their inward
migration is sufficiently rapid that they do not acquire a significant
amount of gas along the way \citep{Ivanov}.  Termination of migration
allows low-mass planets (formed at $a \la 0.7$AU in
relatively massive disks) to grow through gas accretion to their
asymptotic mass $M_{\rm p,trunc} \simeq 30 M_\oplus$ at the
stalling location which is set to be $a_{\rm stall} =0.04$ AU.  
Thus, the final mass of close-in planets with 
$a_{\rm ini} \la 0.7$AU is $\simeq 30 M_\oplus$ regardless of 
truncation mass $M_{\rm p,trunc}$ at the original locations $a_{\rm ini}$.
These planets are formed interior to the ice boundary and they are likely to be
mostly composed of silicates and iron, in contrast to the ice giants
in the solar system.  Also, they cannot accrete large amount of gas
because the aspect ratio of their nascent disk at 0.04 AU is so small
that they induce clear gap formation when their mass reaches that of
Neptune.

Planets formed at slightly larger $a$'s ($\sim 1$AU) must 
attain $M_{\rm p} \ga 20-30 M_\oplus$ 
before they acquire the mass to start migration, 
$M_{\rm p,mig} = 10 M_{\rm g,vis}$ 
(eq.~[\ref{eq:m_gas_vis}]).
With this critical mass, embryos which can initiate migration there can
also accrete gas efficiently with $\tau_{\rm KH} < \tau_{\rm mig}$
provided the gas in the disk is not severely depleted.
Since $M_{\rm p,trunc}$ decreases with decrease in $a$, 
when they arrive close to their host stars, 
their $M_{\rm p}$ may be $> M_{\rm p,trunc}$ so
that they would not acquire any additional mass.  Note that cores of
these planets are also mostly made of silicates and iron rather than
ice and they should not be referred to as hot Neptunes.  

At even larger radius, $M_{\rm c, iso} \ga 5-10 M_\oplus$. 
Prior to reaching isolation, embryos' gas accretion is 
suppressed by the bombardment of residual
planetesimals, {\it i.e.,} $M_{\rm c} < M_{\rm c, hydro}$ for
moderate $\dot{M}_{\rm c}$.
After isolation is reached, 
$M_{\rm c, hydro}$ becomes smaller than $M_{\rm c}$ and
their $\tau_{\rm KH}$
due to gas accretion is reduced below $\tau_{\rm mig}$ and they
quickly evolve into gas giants.  
Thus, final $M_{\rm p}$ in Figure \ref{fig:ma}b coincides with
$M_{\rm p,trunc}$ at initial locations.
Beyond $\sim 10$ AU, the time scale
for the emergence of cores with $M_{\rm p} > M_{\rm c,acc}$ or
isolation is comparable to or longer than both $\tau_{\rm mig}$ or
$\tau_{\rm dep}$.  Although they may acquire $M_{\rm p}$ in the range of
$10-100 M_\oplus$ through merges of residual embryos after the gas 
depletion, these planets generally do not migrate extensively.  

\subsection{Mass distribution of short-period planets}

Interior to the ice boundary, nearly all the planets with sufficient mass
to initiate efficient gas accretion have migrated to the vicinity of
the host star (see Figure \ref{fig:ma}b). But, a majority of the planets which
migrated to the vicinity of solar-type stars were formed beyond the
ice boundary, as gas giants, prior to their migration.  There is a narrow
window in the range of $a$ where the seed of intermediate-mass planets
may form and migrate to the proximity of their host stars.  
In Figure \ref{fig:histo},
theoretically predicted mass distributions are plotted.
Since it is expected that most of close-in planets may
fall onto their host stars, we plot the distributions
of close-in planets, reducing the amplitude $N$ by a factor 10.
For comparison of the amplitude between close-in and distant planets,
uncertainty in this calibration is noted.
We also plotted observed distributions.
Since the number of runs in each model does not 
reflect the number of targets for current doppler survey,
we cannot compare the amplitude $N$ between the observed and 
the theoretically predicted distributions.
Only the shape of the distributions should be compared.
Also note that observed distributions do not exactly
correspond to host stars' mass of each model and
numbers of observed planets are not large enough for
statistical arguments for stars other than G stars (model 1.0).  
We show that, around stars with $M_\ast \simeq 1 M_\odot$, the mass
distribution for the close-in (with $a_{\rm fin} < 0.05$ AU) planets
is skewed toward $\sim 10^3 M_\oplus$ (model 1.0 
in Figure \ref{fig:histo}a). 
This distribution is more enhanced near $\sim 10^3 M_\oplus$ than
that observed.  The effect of post
formation star-planet tidal interaction, which has not been 
taken into account in our model, may have caused the demise of a majority of
the close-in planets, in particular massive planets 
(Gu {\it et al.} 2003, Paper II).

In disks with modest masses, planets form in the advance stages of
evolution when gas depletion is well underway.  Some of these planets
may migrate interior to the ice boundary and become stalled while others
remain close to their place of birth.  A majority of gas giant planets 
with $0.1 {\rm AU} \la a_{\rm fin} \la 1$ AU 
have migrated but not extensively.
In contrast to the
close-in planets, 
model 1.0 in Figure \ref{fig:histo}b clearly shows a paucity of 
longer-period ($0.1 {\rm AU} \la a_{\rm fin} \la 1$ AU) planets at 
$M_{\rm p} \sim 20-100M_\oplus$.
This distribution reflects the stringent prerequisite that gas accretion into
gas giants must be preceded by the rapid formation of sufficient
mass cores whereas the build up of terrestrial planets can continue
well after the severe depletion of the disk.

\section{Dependence on the stellar mass}

In a generalization of the solar nebula model, Nakano (1988) showed
that the temperature distribution throughout the disk increases with
$M_\ast$.  However, he did not consider the dependence of $\Sigma_{\rm
d}$ and $\Sigma_{\rm g}$ on $M_\ast$.  In a recent paper, 
\citet{Laughlin04b}, considered a model in which the disk mass
increases with the stellar mass.  Their objective is to demonstrate
the difficulties to form gas giants around M dwarf stars.  But, the
effects of planetary migration, truncation of gas accretion due to gap opening 
and the gradual depletion of the disk gas are neglected.

In this section, we consider the variation of 3 model parameters: 1) the
stellar mass $M_\ast$, 2) the dependence of disk mass 
on the stellar masses, $h_{\rm d}$, 
and 3) the condition for the onset of type II migration.
A standard series are Model $x$ where
$x = 0.2, 0.4, 0.6, 1.0, 1.5$ represents 
models with $M_\ast = x M_\odot$.
In the standard series, 
we set $A_\nu =10$ and $h_{\rm d} = (M_\ast/M_\odot)^{2}$.
Model $x$B (series B) and $x$C (series C) correspond to 
$A_{\nu} = 1$ and 100 with $h_{\rm d} = (M_\ast/M_\odot)^2$.
$x$ in $x$B and $x$C expresses $M_\ast/M_\odot$ as well.
Model $x$D (series D) and $x$E (series E) correspond to 
$h_{\rm d} = M_\ast/M_\odot$ and 1 with $A_{\nu} = 10$.

In Figure \ref{fig:f_d_dist}, 
distributions of $f_{\rm d}h_{\rm d}$ we used for 
models $x$ and $x$B are shown.
They represent the relative mass
distribution of disks around stars with various masses. The mean value
of $\Sigma_{\rm d}$ is an increasing function of $M_{\ast}$ 
(see eq.~[\ref{eq:sigma_dust}]).
In order to limit additional model
parameters, we assume ZAMS mass-luminosity relationship, 
$L_\ast/L_\odot = (M_\ast / M_\odot)^4$.  
Since the time scale of pre-main sequence stage of lower-mass stars
is long, planet formation around these stars may proceed during their 
pre-main sequence stage in which $L_\ast$ is rather large.
In that case, the dependence of $L_\ast$ on $M_\ast$ is weaker, but
it may still have a positive power-law dependence, so that 
the trend of the $M_\ast$-dependence of planetary systems 
shown below does not change.

In the calculations in this section,
\begin{equation}
a_{\rm ice} = 2.7 \left( \frac{M_\ast}{M_\odot} \right)^2 {\rm AU},
\label{eq:aice_sp}
\end{equation}
\begin{equation}
M_{\rm p}(t) =
\left( \frac{t}{4.8 \times 10^5{\rm years}} \right)^3 \eta_{\rm ice}^{3}
f_{\rm d}^{21/5} h_{\rm g}^{6/5}
\left( \frac{a}{1{\rm AU}} \right)^{-81/10}
\left(\frac{M_\ast}{M_{\odot}} \right)^{89/10}
M_{\oplus},
\label{eq:m_grow0_sp}
\end{equation}
\begin{equation}
M_{\rm c,iso} =
0.16 \eta_{\rm ice}^{3/2} f_{\rm d}^{3/2} 
\left(\frac{a}{1\mbox{AU}}\right)^{3/4} 
\left(\frac{M_\ast}{M_{\odot}} \right)^{5/2} M_{\oplus},
\label{eq:m_iso_sp}
\end{equation}
\begin{equation}
M_{\rm c,no iso} \simeq
1.2 \eta_{\rm ice} f_{\rm d}
\left(\frac{a}{1\mbox{AU}}\right)^{1/2} 
\left(\frac{M_\ast}{M_{\odot}} \right)^{2} M_{\oplus}, 
\label{eq:m_noiso_sp}
\end{equation}
\begin{equation}
M_{\rm e,iso} =
0.52 \eta_{\rm ice}^{3/2} f_{\rm d}^{3/2} 
\left(\frac{a}{1{\rm AU}}\right)^{3/2}
\left(\frac{\rho_{\rm d}}{1{\rm gcm}^{-3}}\right)^{1/4} 
\left(\frac{M_\ast}{M_\odot}\right)^{9/4} M_{\oplus},
\label{eq:m_e_iso_sp}
\end{equation}
\begin{equation}
M_{\rm p,mig} = A_\nu M_{\rm g,vis} = 30 \left(\frac{A_\nu}{10}\right)
                      \left(\frac{\alpha}{10^{-4}}\right)
                      \left(\frac{a}{1{\rm AU}}\right)^{1/2}
                      \left(\frac{M_\ast}{M_{\odot}}\right) M_{\oplus},
\label{eq:m_gas_vis_sp}
\end{equation}
and
\begin{equation}
M_{\rm p,trunc} = A_{\rm th} M_{\rm g,th} = 400 \left(\frac{A_{\rm th}}{1.5^3}\right)
\left(\frac{a}{1{\rm AU}}\right)^{3/4}
\left(\frac{M_\ast}{M_\odot}\right) M_\oplus.
\label{eq:m_gas_th_sp}
\end{equation}

\subsection{Planets around M dwarf stars}

\begin{figure}
\plotone{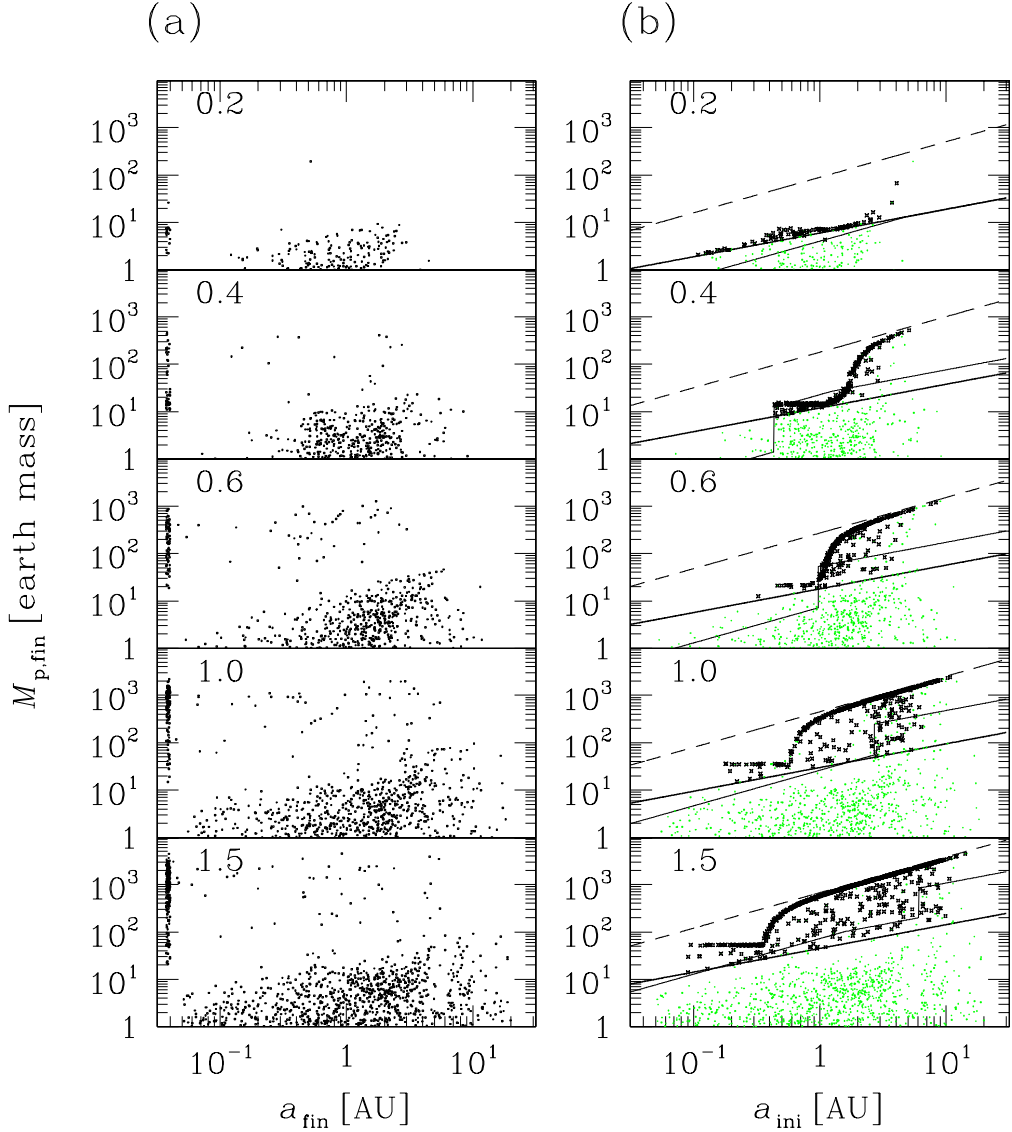}
\caption{
The distributions of semimajor axis ($a$) 
and mass ($M_{\rm p}$) of planets predicted by the Monte Carlo simulations.
(a) final semimajor axes ($a_{\rm fin}$) and masses ($M_{\rm p,fin}$) 
at $t = 10^9$ yrs, and (b) 
initial semimajor axes ($a_{\rm ini}$) and $M_{\rm p,fin}$.
The symbols $x$ ($= 0.2, 0.4, 0.6, 1.0$ and 1.5)
represent the host star mass scaled by the solar mass, $M_{\ast}/M_\odot$.  
In (b), close-in planets with $a_{\rm fin} < 0.05$AU are marked by black 
crosses,
while the other planets are marked by gray dots.
The thin solid black lines indicate
the isolation mass $M_{\rm c, iso}$ with $f_{\rm d} =30$.
The thick solid black lines express the critical mass for 
radial migration, $M_{\rm p, mig} = A_\nu M_{\rm g, vis}$
with $A_\nu = 10$.
The dashed lines are the truncation mass for 
gas accretion, $M_{\rm p, trunc} = A_{\rm th} M_{\rm g, th}$
with $A_{\rm th} = 3.4$.
}
\label{fig:ma}
\end{figure}

\begin{figure}
\plotone{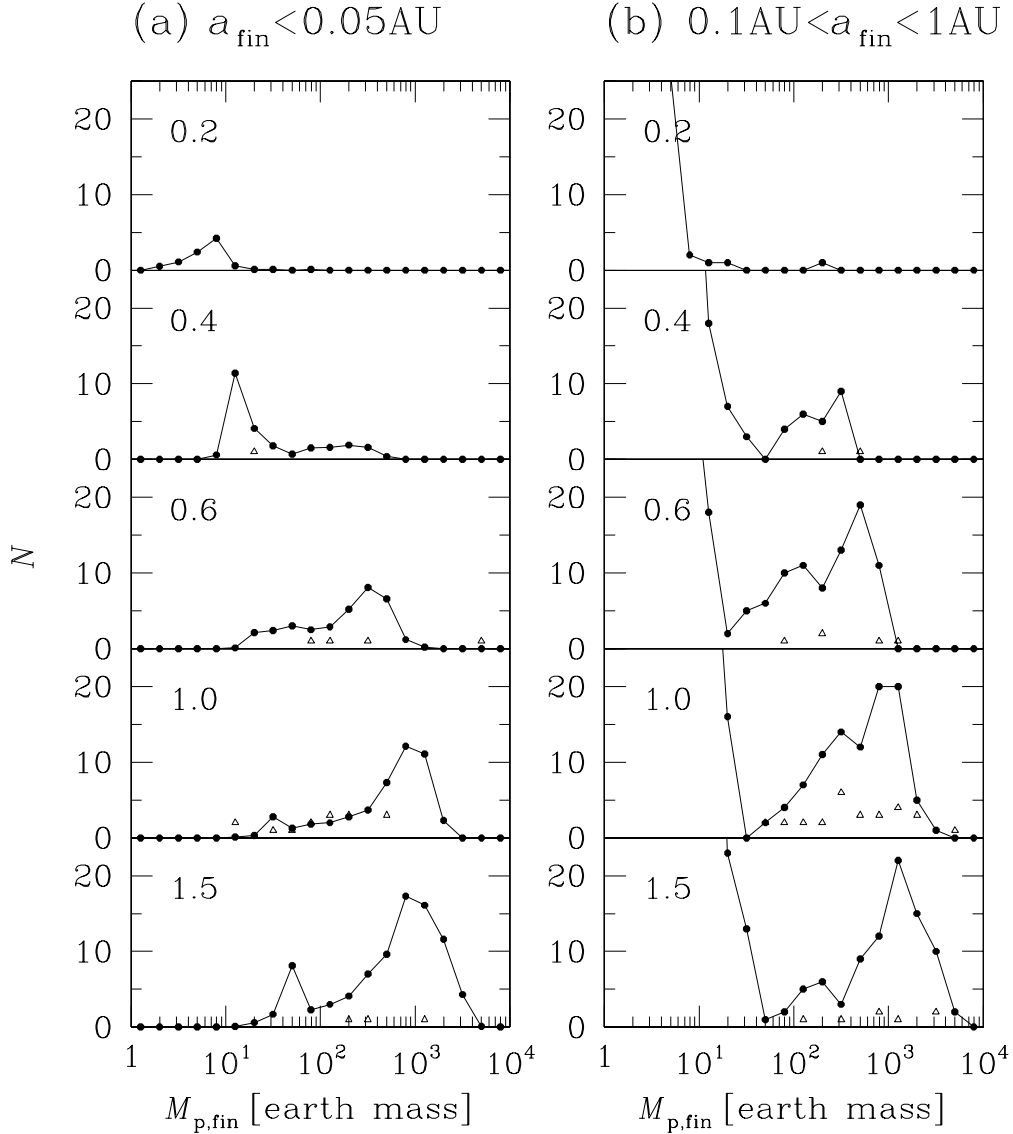}
\caption{
The calculated distribution of final mass of planets for
(a) close-in planets at $a_{\rm fin} < 0.05$AU and
(b) planets at 0.1AU $< a_{\rm fin} < 1$AU (filled circles).
The labels $x$ are the same as Figure \ref{fig:ma}.
Since it is expected that most of close-in planets may
fall onto their host stars, the calculated amplitude $N$ 
in (a) is reduced  by a factor 10.
Observed distributions are also plotted with open triangles.
The $\sin i$ factor is neglected for simplicity
(it enhances the observed values only by $4/\pi$ on average). 
The number of runs in each model does not 
reflect the number of targets for current doppler survey.
Also note that observed distributions do not exactly
correspond to host stars' mass of each model.
}
\label{fig:histo}
\end{figure}

\begin{figure}
\plotone{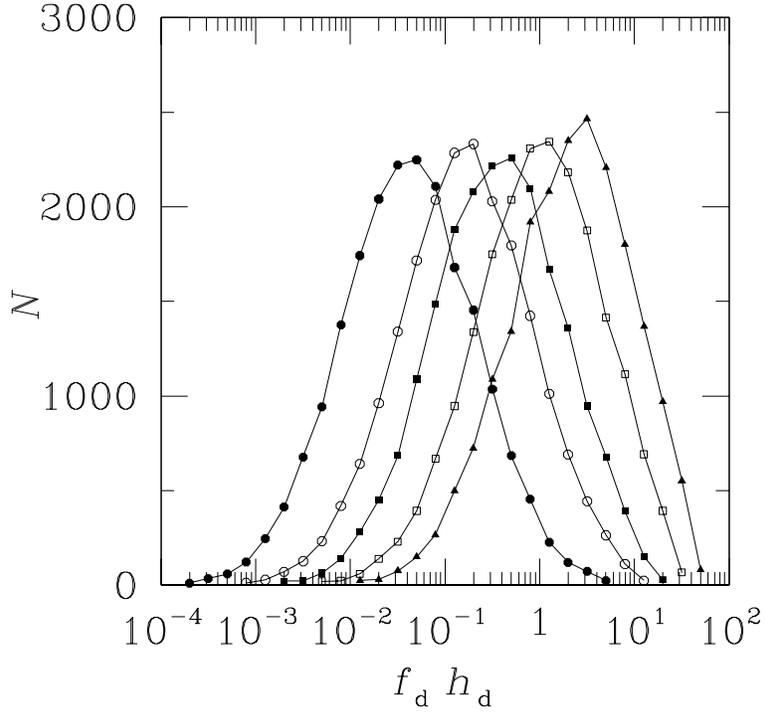}
\caption{
The $f_{\rm d} h_{\rm d}$ distribution we used for
the standard model, which
is a gaussian distribution in terms of $\log_{10} f_{\rm d}$
with a center at $\log_{10} f_{\rm d} = 0$ and dispersion of 1. 
In the standard model, we assume $h_{\rm d} \propto (M_\ast/M_\odot)^2$.
We omit the high $h_{\rm d} f_{\rm d}$ tail at $ > 30 (M_\ast/M_\odot)$, 
since such heavy disks are self gravitationally unstable. 
Filled circles, open circles, filled squares, open squares
and filled triangles represent the cases of
$M_{\ast} =$ 0.2, 0.4, 0.6, 1.0, and 1.5$M_{\odot}$.
}
\label{fig:f_d_dist}
\end{figure}

In Figures \ref{fig:ma} and \ref{fig:histo}, 
we also included the results of the simulations
for models 0.2-0.6 and 1.5.  
We first present a low-mass model 0.2 since it is in
strong contrast to model 1.0. 
Stars with $M_\ast = 0.2 M_\odot$ corresponds 
to relatively light M stars.
These stars are not only most numerous but they also contribute 
most to the initial
stellar mass function.  According to our prescription and model
parameters, $\Sigma_{\rm g}$ and $\Sigma_{\rm d}$ around the host star
in model 0.2 are 25 times smaller than those in model 1.0.

In the result of model 0.2 on the top panels of 
Figures \ref{fig:ma}a and b, 
we find that Jupiter-mass
planets rarely formed around low-mass stars. This paucity
is due to the slow growth rate of embryos and their low isolation mass
such that little gas can be attained by them prior to its depletion.
This result confirms the conclusion reached earlier by \citet{Laughlin04b}.

The upper limit of the $M_{\rm p, fin}-a_{\rm fin}$ distribution
is determined by the $M_{\rm c, iso}$.  This correlation arises
because $L_\ast$ is a rapidly rising function of $M_{\ast}$.  In our
prescription, the ice boundary is located at $a_{\rm ice} \simeq 0.11$AU
(eq.~[\ref{eq:aice_sp}]) which is 25 times closer to
a host star with $M_\ast = 0.2 M_\odot$ than in model 1.0 with
solar-type stars.  Nearly all the cores formed around these low-mass
stars are mostly composed of ice.  Planets can acquire masses $M_{\rm
p} > M_{\rm c, iso}$ through gas accretion.  But with a relatively
small $M_{\rm c, iso}$, the accretion is inefficient 
(eq.~[\ref{eq:tau_KH}]). They can
also gain mass after the gas depletion through collisions and mergers
of residual embryos. But the asymptotic mass $M_{\rm e,iso}$ is also 
relatively small for low-mass stars (eq.~[\ref{eq:m_e_iso_sp}]).

With the {\it ad hoc} $\alpha$ prescription we have adopted, the
necessary condition for the onset of type II migration is satisfied
for relatively low-mass planets
(eq.~[\ref{eq:m_gas_vis_sp}]). When their $M_{\rm p} > M_{\rm p, mig}$, 
these cores undergo orbital decay.  Similar to model 1.0,
there is a population of close-in planets with $2 M_\oplus \la M_{\rm p}
\la 10 M_\oplus$.  The results on the top panel of Figure \ref{fig:ma}b show that
they indeed originated from a region between $a_{\rm ice}$ and $\sim
3$ AU. They also indicate that planets with mass down to $2 M_\oplus$
may migrate to the proximity of a $0.2 M_\odot$ star.  At the
arbitrary $a_{\rm stall} = 0.04$ AU around such a host star, the
equilibrium temperature of these close-in Neptune-mass planets is
similar to that of the Earth. The composition and structure of such
planets have been already been discussed by \citet{Leger04}.

Although a few planets formed with $M_{\rm p} > 10 M_\oplus$, they are
the exceptional cases resulting from the tails of the $f_{\rm d}$
distribution. Consequently, the mass function of planets with modest
or large $a_{\rm fin}$ ($> 0.1$ AU) show a sharp decline at $M_{\rm
p} \sim 10 M_\oplus$ (Figure \ref{fig:histo}b).  It also extends well into the low-mass
range. In contrast, the mass function for the close-in planets (with
$a_{\rm fin} < 0.05$ AU) shows a peak near $M_{\rm p} \sim 10
M_\oplus$ (Figure \ref{fig:histo}a). 

In model 0.4, we consider a host star with $M_\ast = 0.4 M_\odot$ which
corresponds to a relatively massive M dwarf star.  
It is an analog of GJ436, around
which a single close-in Neptune-mass planet has been discovered
\citep{Butler04}. 
Figure \ref{fig:histo}a and b indicate that 
the frequency of planets observable with current doppler
survey is significantly smaller for M stars than for
K, G, and F stars (also see Figure \ref{fig:freq}).
This is consistent with the observation and 
the finding of \citet{Laughlin04b}.

\begin{figure}
\plotone{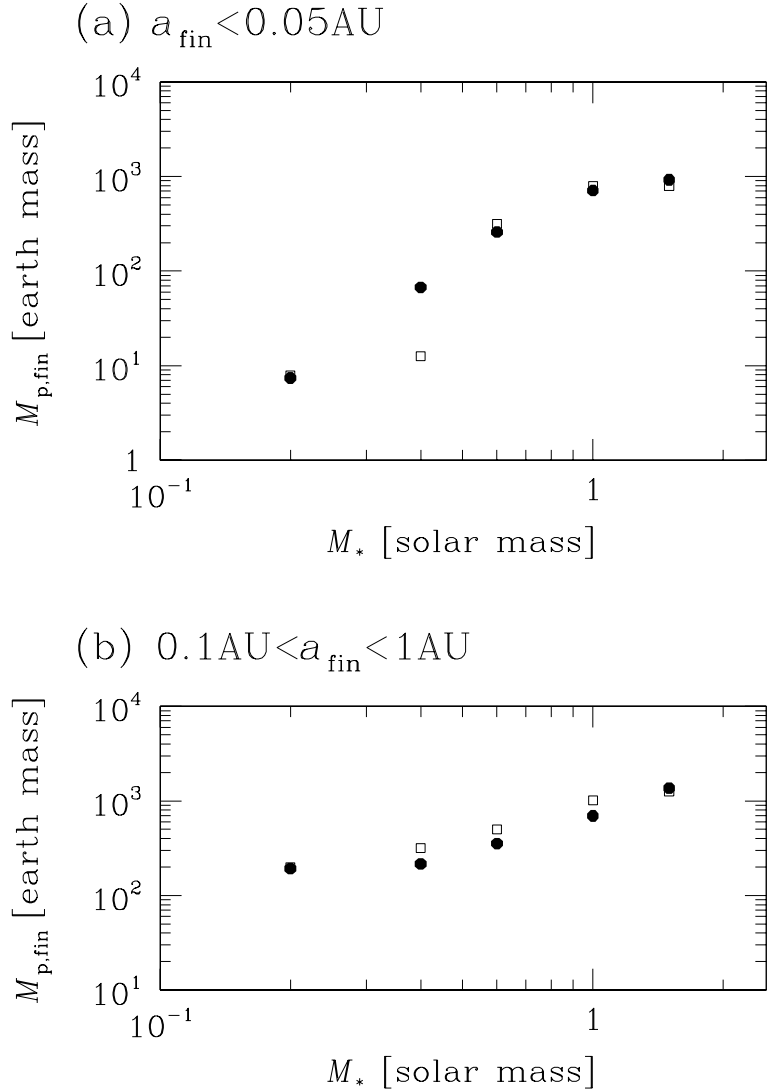}
\caption{
The mean mass and the characteristic mass associated with the peak
of the mass distribution given in Figures \ref{fig:histo},
as a function of host stars' mass $M_\ast$.
(a) the close-in planets with $a_{\rm fin} < 0.05$AU and
(b) the planets at 0.1AU $< a_{\rm fin} < 1$AU.
The mean mass and the peak mass are plotted with
filled circles and open squares.
For (b), only planets with $M_{\rm p}$ over a deficit
($M_{\rm p} > 50 M_\oplus$) are considered 
(see Figures \ref{fig:ma}a and \ref{fig:histo}b).
}
\label{fig:mean_mp}
\end{figure}

Figure \ref{fig:mean_mp} indicates 
the mean mass and characteristic mass associated with the 
highest peak of the mass distribution for a) the close-in planets and
b) planets with $0.1 < a < 1$AU.
These masses increase with $M_\ast$, 
in particular for the close-in planets.
This accounts for the recent planetary finding around GJ436.

\begin{figure}
\plotone{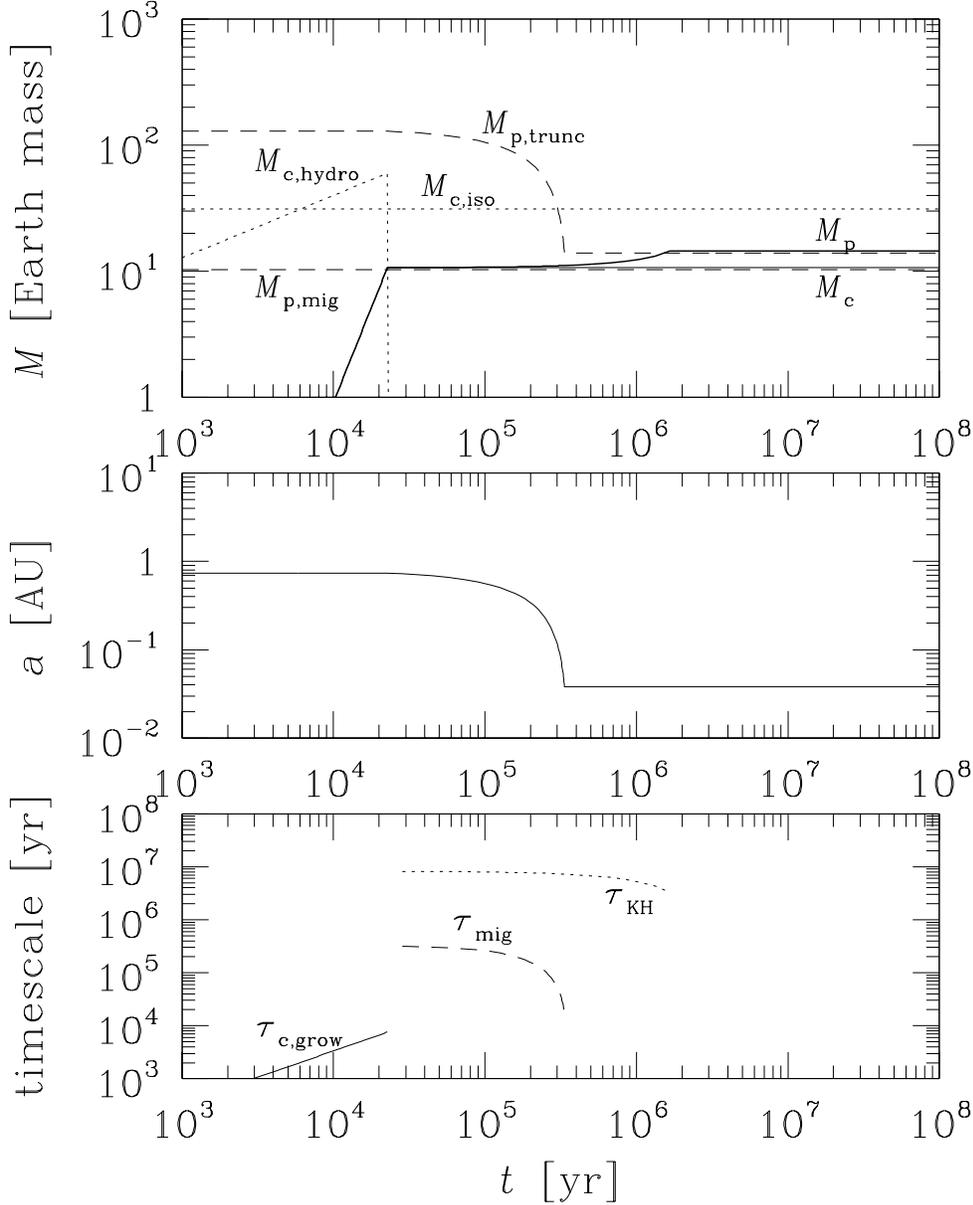}
\caption{
An example of formation of a close-in Neptune-mass planet around an 
M star.
A plant is initially at 0.74AU in a disk with $f_{\rm d} h_{\rm d} = 7.2$
and $\tau_{\rm dep} = $ 9.2 Myr around a star with $M_\ast = 0.4 M_\odot$.
The upper and middle panels show
time evolution of mass and semimajor axis of the planet.
In the upper panel,
the isolation mass of a core ($M_{\rm c,iso}$),
its critical mass for initiation of gas accretion ($M_{\rm c,hydro}$),
the planet's critical mass for migration ($M_{\rm g,mig}$) and
the truncation mass of gas accretion ($M_{\rm g,trunc}$) are plotted
for comparison.
In the lower panel,
core accretion time scale ($\tau_{\rm c,grow} = M_{\rm c} / \dot{M}_{\rm c}$),
gas accretion time scale ($\tau_{\rm KH}$) and
migration time scale ($\tau_{\rm mig}$) are plotted.
}
\label{fig:tevol_mstar04}
\end{figure}

We find {\it the frequency of Neptune-mass close-in planet peaks at
$M_\ast \sim 0.4 M_\odot$} (Figures \ref{fig:histo}a and
\ref{fig:mean_mp}).  The formation of these
planets depends sensitively on the environment and their frequency
provides constraints on the sequential accretion scenario.  Most of
these planets formed slightly beyond the ice boundary.  For this 
stellar mass, the ice boundary is located at $\simeq 0.43$ AU.  Outside
it, the upper limit of $M_{\rm c, iso}$ exceeds $M_{\rm p, mig}$.  
In Figure \ref{fig:tevol_mstar04}, 
we illustrate the evolution of a typical embryo
that forms near the ice boundary in a disk with 
$h_{\rm d} f_{\rm d} =7.2$ and a depletion time 
scale $\tau_{\rm dep} = 9.2$ Myr.
The disk surface density $h_{\rm d} f_{\rm d} =7.2$
is close to the tail of the distribution for $0.4 M_\odot$ stars
(Figure \ref{fig:f_d_dist}).
During the initial $2 \times 10^4$ yrs, an embryo grows to 
$\sim 10 M_\oplus$ through coagulation.
The rapid coagulation enhances $M_{\rm c,hydro}$, which
prevents gas accretion (see eq.~[\ref{eq:crit_core_mass}] and 
Figure \ref{fig:tevol_mstar04}).
Since this embryo is located beyond the ice boundary
and $h_{\rm d} f_{\rm d} =7.2$, $M_{\rm c,iso}$ is as large 
as $\simeq 30M_\oplus$.  But, before it acquires $M_{\rm c,iso}$,
it opens up a gap and 
undergoes type II migration at $M_{\rm c} \sim 10M_\oplus$.
During the migration, gravitational perturbations from the embryo
prevents additional planetesimals from reaching the embryo 
and its growth is quenched  \citep{TI97, Rafikov03}.  On
a time scale of $3 \times 10^5$ yr, the embryo migrates to 0.04 AU
where it is stalled.  
The termination of bombardment by residual planetesimals 
makes $M_{\rm c,hydro}$ lower to start gas accretion onto the embryo.
However, the gas accretion time scale $\tau_{\rm KH}$ is longer than 
$\tau_{\rm mig}$ (Figure 4).
Gas accretion onto the planet actually proceeds after it
reaches 0.04AU.
The planet's growth is quenched at $\simeq 14M_{\oplus}$ 
as a clean gap is formed because $M_{\rm p,trunc} \simeq 14M_{\oplus}$ 
at 0.04AU.  
Figures \ref{fig:ma} and \ref{fig:histo}
show that many planets evolve in a similar way to
acquire $M_{\rm p,fin} \simeq 14M_{\oplus}$. 
The composition of these planets
is similar to that of Uranus and Neptune and its day-side surface
temperature may be $\sim 500$K.

In the lowest-mass model 0.2, the upper limit of
$M_{\rm c, iso}$ does not exceed $M_{\rm p, mig}$ even outside the
ice boundary.
The migration of Neptune-mass planets
requires a delicate balance between $M_{\rm c, iso}$ and $M_{\rm p, mig}$.
Thus, the frequency of 
close-in Neptune-mass planets in model 0.4 is larger than that in model 0.2.

This frequency in model 0.4 is also larger than that in model 1.0.
Although the upper limit of $M_{\rm c, iso}$ is larger than $M_{\rm p, mig}$
outside the ice boundary around solar-type stars, 
$M_{\rm p} \ga 30-40 M_\oplus$ at the onset of migration, so that
gas accretion is much more
efficient onto such large cores (eq.~[\ref{eq:tau_KH}]).  
Thus, most isolated planets tend to
arrive at the proximity of a solar-type star as gas giants and only a
small fraction arrives as silicate and iron cores with limited gaseous
envelopes.  

Recently Neptune-mass planets have been detected around 
55Cnc and HD160691 \citep{McArthur04, Santos04}.  
Since these stars are G-type stars, if the result of Figure \ref{fig:histo}
is applied, the probability of formation of such planets is very low 
although it is not zero.
Unlike GJ436, there are three additional Jupiter-mass planets
around 55Cnc and two additional ones around HD160691.
In the system of 55Cnc, sweeping mean motion resonance 
associated with migration of the giant planet 
\citep{Malhotra, Ida00} presently at 0.1AU
could bring rocky embryos/planetesimals to the vicinity
of the host star, so that a Neptune-mass rocky planet
could accrete {\it in situ}.
On the other hand, in the system HD160691, 
sweeping secular resonance associated with disk gas depletion
\citep{Ward81, Nagasawa00}, 
could bring rocky embryos/planetesimals to 
inner regions \citep{Lin_etal04}.
In the present paper, we do not include such interactions.
We will present elsewhere details.
In the system of GJ436, however, no additional Jupiter-mass 
planet has been found.
Our model accounts for formation of the isolated Neptune-mass
planet around an M star.

\subsection{Planets around K dwarfs}

Figure \ref{fig:histo}b shows that
in model 0.6 where $M_\ast=0.6 M_\odot$, the intermediate-mass
($20-100M_\oplus$) and intermediate-$a$ ($0.1-1$AU) planets
are more abundant than in either model 1.0 or model 0.4.
The mass function of close-in planets also appears to be
smoother with some $M_{\rm p}$ in the range of $\sim 10-100 M_\oplus$ 
(Figure \ref{fig:histo}a).  
These intermediate-mass planets formed just outside the ice boundary in disks
with modest $f_{\rm d}$ where the upper limit of $M_{\rm c, iso} \sim
50-100 M_\oplus$ while that for typical disks (with $f_{\rm d} \sim 1$) is
$\sim M_\oplus$.  The critical mass for starting planetary
migration at the ice boundary is $20 M_\oplus$ in model 0.6.  A small
fraction of emerged cores may accrete modest amount of gas as it starts
to migrate.

\begin{figure}
\plotone{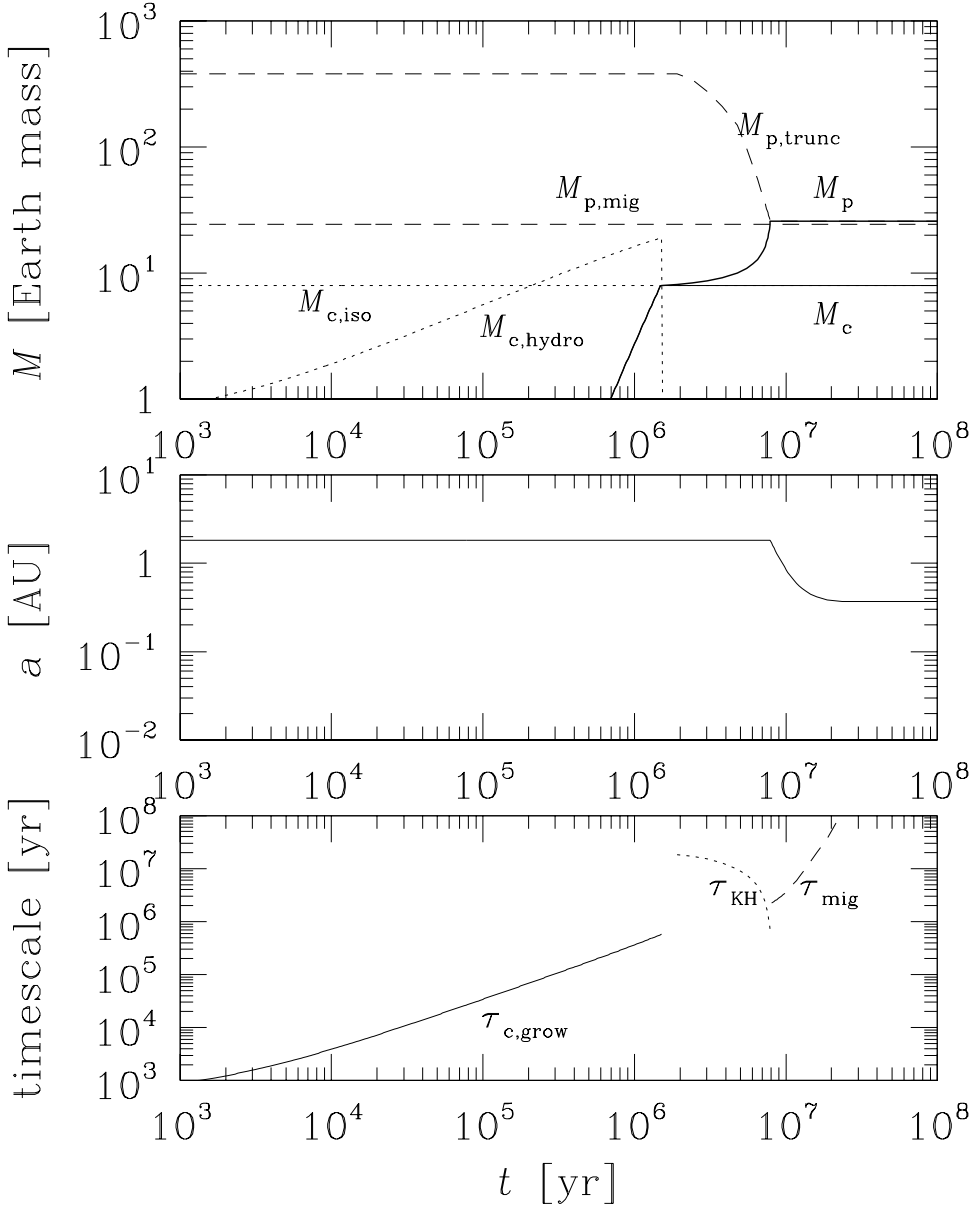}
\caption{
An example of formation of an intermediate-mass planet
with intermediate period around a K star.
A plant is initially at 1.8AU in a disk with $f_{\rm d} h_{\rm d} = 2.0$
and $\tau_{\rm dep} = $ 2.3 Myr around a star with $M_\ast = 0.6 M_\odot$.
Notations are the same as Figure 5.
}
\label{fig:tevol_mstar06}
\end{figure}

The formation of the intermediate-mass
and intermediate-$a$ planets, which tends to smooth 
the mass distribution, may be one of characteristics
of planets around K stars, compared with those around
G and F stars, although it is much less pronounced
than the characteristics of planets around M stars.
For illustration, we present, in Figure \ref{fig:tevol_mstar06}, 
the formation of a typical
planet which formed with an intermediate mass and attain an
intermediate $a$ during its migration.  In this case, a seed embryo
is formed at $a_{\rm ini} =1.8$AU in a slightly massive $h_{\rm d}
f_{\rm d} = 2$ disk with $\tau_{\rm dep} = 2.3$Myr.  Through
planetesimal coagulation, this embryo attains a mass $M_{\rm p} \simeq
8 M_\oplus$ and become isolated in 1.5 Myr.  The cessation of the
planetesimal bombardment enables the embryo to grow through gas
accretion.  When its mass reaches $M_{\rm p} = 30 M_\oplus$ at $\sim 8$
Myr, gas accretion is quenched by the severe depletion of gas near its
orbit.  The newly formed planet undergoes migration while gas is
globally depleted.  The orbital migration is eventually halted at 
an intermediate location, 0.4AU. 

The condition for growth to be quenched between $10-100 M_\oplus$ 
requires $a \sim 1$ AU.  The above example shows that to halt migration at an
intermediate location, both time scales of migration and
growth due to gas accretion are required to be comparable
to the gas depletion time scale.  
In general, such special circumstances are satisfied
with small probability. 
However, they are more likely around K dwarfs than other type stars
because cores with $M_{\rm p} \ga 10 M_\oplus$ are
more abundant at $a \sim 1$AU around K dwarfs 
as a result of the $M_{\ast}$ dependence of
$a_{\rm ice}$ and $M_{\rm c, ico}$.
In model 0.6,
gas accretion may be quenched when the cores attain a mass $10
M_\oplus \la M_{\rm p} \la 100 M_\oplus$ while they migrated to 
$0.04 {\rm AU} \la a \la 1$ AU.  
In comparison
with model 1.0, the isolation mass $M_{\rm c, iso}$ at the ice boundary 
increases with $M_\ast$. 
Around a solar-type star, the isolation mass near the ice
boundary is sufficiently large for efficient gas accretion to be
initiated.  Eventually runaway gas accretion leads to the emergence of
the intermediate-mass deficit in the mass distribution of planets around
relatively high-mass stars.  
Figure \ref{fig:obs} shows that some fraction of
planets discovered around K stars may have 
the intermediate mass.
However, the number of the detected planets may be insufficient
for statistical discussion.
These stellar mass dependence in the
extrapolated planetary characteristics can be tested with future
observation.

\subsection{Planets around higher-mass stars}

\begin{figure}
\plotone{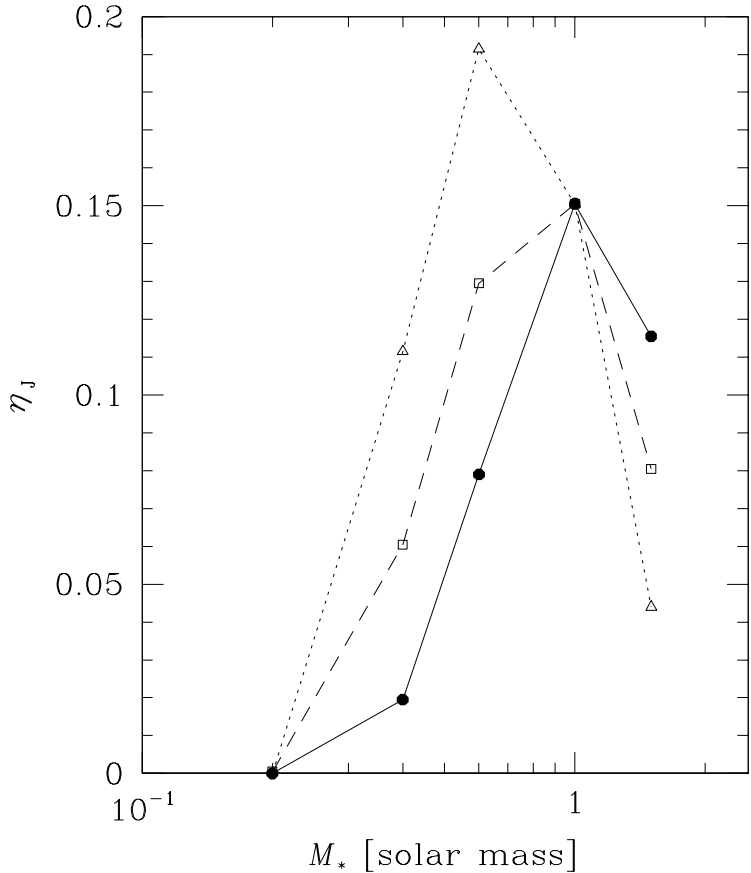}
\caption{
The theoretically predicted 
fraction ($\eta_{\rm J}$) of stars which bears giant planets 
currently detectable with current doppler survey, 
as a function of their mass ($M_\ast$). 
The results of the standard model with
$h_{\rm d} = (M_{\ast}/M_\odot)^2$ are plotted by filled circles,
while the results of series D and E with
$h_{\rm d} = M_{\ast}/M_\odot$ and 1 are plotted by 
open squares and triangles, respectively.
The detectable conditions are radial velocity $v_r > 10$m/s
and period $P < 4$ years.
We do not include planets stalled at 0.04AU in the evaluation of
$\eta_{\rm J}$, since most of them may fall onto their host stars.
The distribution of $f_{\rm d}$ is the same as in Figure 
\ref{fig:f_d_dist}.  For
each $f_{\rm d}$, $a$ is selected as $\log (a_{j+1}/a_j) = 0.2$
$(j=1,2,...)$.  
}
\label{fig:freq}
\end{figure}

In model 1.5 where $M_\ast=1.5 M_\odot$, 
the range of $a$ where gas giants are formed is more extended
(the bottom panel in Figures \ref{fig:ma}). 
This arises primarily because the disks around more massive stars
have relatively large $\Sigma_{\rm d}$. 
In this case, $M_{\rm c, acc}$ (the core mass required for 
rapid gas accretion) can be attained before gas depletion even 
at large $a$.
However, larger $a_{\rm ice}$ in this case leads
to slow core growth beyond the ice boundary.   Thus,
most gas giants have cores composed of silicates and iron
but not icy cores, in contrast with gas giants around lower mass stars.
The very large $a_{\rm ice}$ also leads to less efficiency 
of formation of gas giants in the range of a few AU to 10AU
than that around G stars, although
gas giants form in broader range of $a$.
As a result, the fraction of F stars
harboring giant planets with periods smaller than several years
which are currently detectable with doppler survey is
rather smaller than that of G stars (Figure \ref{fig:freq}),
although the difference is within a factor 1.5.
The similar fraction within a factor 2 among K, G, and F stars 
shown in Figure \ref{fig:freq}
is consistent with the observation (Fischer \& Valenti 2005).
The predicted mass distributions for M, K, G, and  F stars
are not inconsistent with the observed distributions.
Since the numbers of observed planets are not enough
for statistical discussion, in particular for stars other than
G stars (Figure \ref{fig:histo}), we cannot discuss the agreement
between the predicted and observed distributions in more detail.

Strong winds and jets from further higher mass stars (A, B stars)
may decrease $\tau_{\rm dep}$,
which reduces formation rate of gas giants.
It is not clear how much the fraction of massive stars with
gas giant planets is reduced.
We will address planetary formation around massive stars elsewhere.

\section{Dependence on the migration condition}

The results of the standard models clearly indicate that the mass
function of planets depends on the delicate balance between 
growth and migration time scales.  
There are some uncertainties concerning the migration
process.  In the standard series, we set $A_\nu =10$ 
(eq.~[\ref{eq:m_gas_vis_sp}) in accordance
with the results of previous numerical simulations \citep{LP85}.
However, the additional contribution from a torque imbalance
between the Lindblad resonances of low-mass embedded cores may also
lead to type I migration while their masses are relatively small
\citep{GT80, Ward86, Ward97},
although turbulence in the disk \citep{Nelson04, Laughlin04a} 
and self induced secondary instability \citep{Balmforth01,
Koller03} can also retard the rate of migration.
In our simulations, we can partly
take into account the effect of type I migration by lowering 
the value of $A_\nu$ (to unity).    
We also carried out simulations with
$A_\nu$ ($= 100$) such that
the onset of migration is delayed until the cores have attained
relatively large masses.

\begin{figure}
\plotone{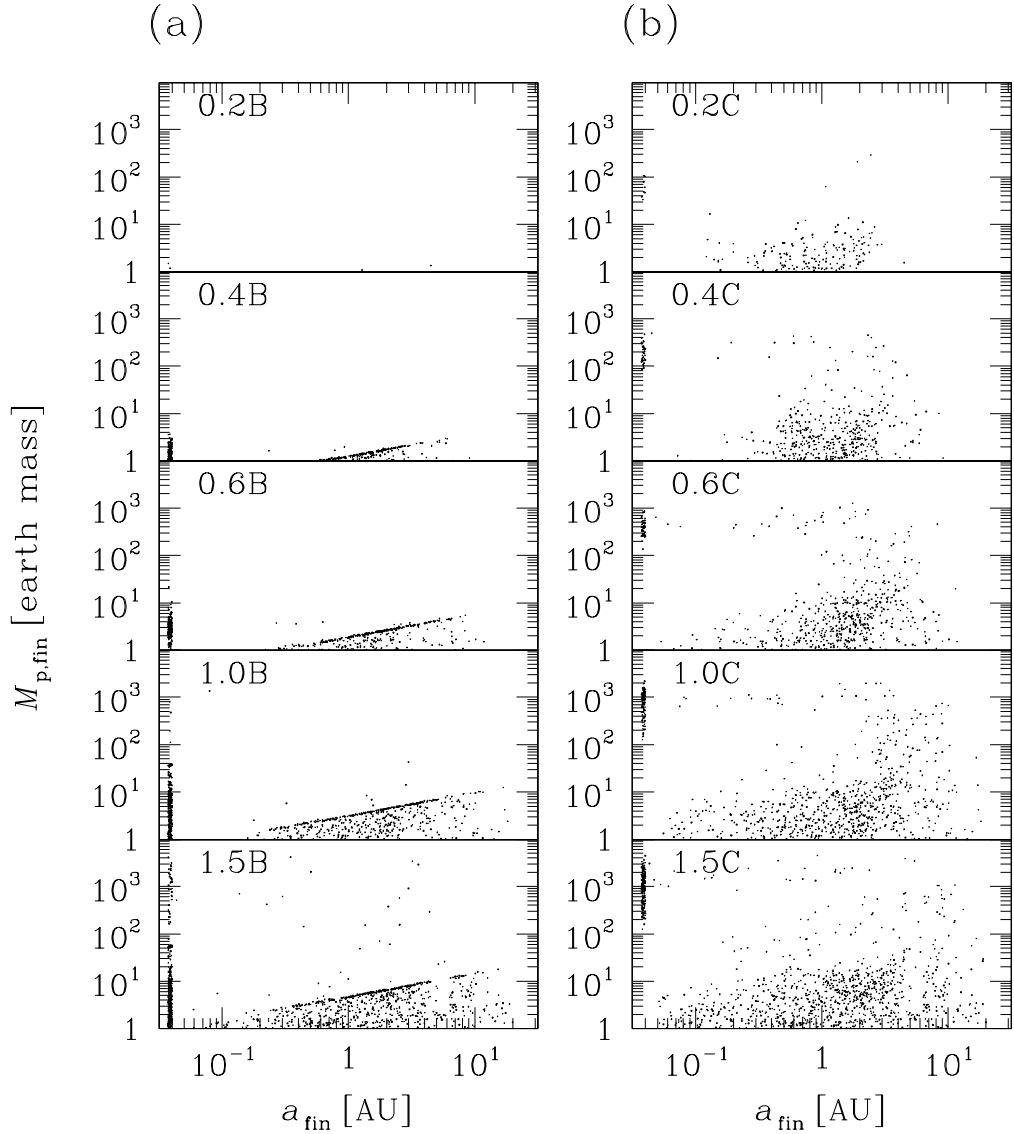}
\caption{
The predicted distributions of semimajor axis ($a$) 
and mass ($M_{\rm p}$) of planets in the final state
at $t = 10^9$ yrs: 
(a) model $x$B with $M_{\rm p, mig} = M_{\rm g, vis}$,
(b) model $x$C with $M_{\rm p, mig} = 100M_{\rm g, vis}$, where
$x$ in $x$B and $x$C represent the results with $M_{\ast} = x M_\odot$.  
}
\label{fig:maBC}
\end{figure}

The sensitive dependence of the mass function of close-in planets on
the migration condition makes it an ideal observable feature which can
be used to calibrate the criteria and efficiency of migration.  In a
variation of the standard models, we consider two new series:
1) series B (model $x$B) with $A_\nu = 1$ and 
2) series C (model $x$C) with $A_\nu =100$, which have
identical $M_\ast$ as the standard series with $A_\nu =10$.
In Figures \ref{fig:maBC}a and b, 
we show the final mass and semimajor axis 
distribution of planets in models $x$B and $x$C.

For models $x$B, cores undergo migration before they attain
sufficient mass to engage in efficient gas accretion.  Although the
cores' migration may terminate close to their host stars, the
relatively small aspect ratio of the disk for small $a$ implies 
low $M_{\rm p,trunc}$.
At small $a$, gap formation prevents the cores from accreting gas.
The accumulation of cores and planetesimals in the proximity of
their host star may promote their coagulation \citep{Ward97b}.  
Although we cannot rule out the possibility of a highly-efficient
migration on the basis of the mass function of the close-in planets,
it does pose difficulties to account for the modest frequency of gas
giants with periods longer than a few weeks.  Even around solar-type
stars, cores rapidly migrate to the stellar proximity before they have
acquired sufficient mass to efficiently accrete gas so that the
probability of gas-giant formation is strongly suppressed.  

For the low-stellar-mass models 0.2B and 0.4B, the isolation mass is only
a few times larger than that of the Earth.  Nevertheless, it is
larger than $M_{\rm p, mig}$ for this low-$A_\nu$ case.  Most
cores migrate toward their host stars with $M_{\rm p}$ less than 
a few $M_\oplus$.  
Although these masses are $ < M_{\rm p, trunc}$, gas
accretion is too slow (eq.~[\ref{eq:tau_KH}]) for them to acquire 
any significant amount of
mass prior to gas depletion.  The detection of close-in Neptune-mass
around M dwarf stars and the modest detection frequency of gas giants
around solar-type stars are inconsistent with the results of series B 
in Figure \ref{fig:maBC}a
and therefore, we suggest that $A_\nu$ is substantially larger than unity.

An upper limit on the magnitude of $A_\nu$ may be inferred from models
$x$C.  With $A_\nu = 100$, the condition for the onset of migration
becomes much more stringent and most cores do not have sufficient
mass to undergo migration.  For solar-type stars, a few relatively
massive cores can form rapidly in disks with very large
$f_{\rm d}$. These systems can migrate to form close-in Jupiter-mass
gas giants.  The mass distribution of
close-in planets around solar type stars (model 1.0C) is skewed to
$10^3 M_\oplus$ with lower cut-off below $\sim 0.5 M_{\rm J}$,
which is inconsistent with the observed mass-period distribution of extra
solar planets (Figure \ref{fig:obs}).  
For low-mass host stars (models 0.2C and 0.4C), many intermediate-mass
planets can form during and after gas depletion.  But they retain
their initial semi major axis.  The mass distribution of the close-in
planets peaks near the mass of Saturn and hardly any planets have
masses comparable to that of Neptune.  These simulation results are
again inconsistent with the observed mass-period distribution of extra
solar planets.  
Furthermore, in series C, the deficit of planets with intermediate masses 
and periods is
too pronounced to be consistent with observed one.
Therefore, we infer $A_\nu \sim 10$.  

\section{Dependence on the disk mass}

The expressions in equations (\ref{eq:m_grow}) and (\ref{eq:m_iso})
indicate that the growth rate and asymptotic mass of cores are
increasing function of $\Sigma_{\rm d}$.  In the standard series of models, we
set $h_{\rm d} = (M_\ast/M_\odot)^2$. With this prescription,
$\Sigma_{\rm d}$ of disks around low-mass stars is relatively small.
Consequently, the emergence of gas giants occurs preferentially around
massive stars.

The dependence of the disks' $\Sigma_{\rm d}$ on the $M_\ast$ of their
host stars is poorly known.  On the theoretical side, gravitational
instability may limit the amount of mass which can be retained by the
disks, especially those around low-mass stars.  But, smaller
$L_\ast$'s and lower intensity of ionizing photons may also reduce the
influence of the magneto-rotational instability \citep{Gammie} and the
angular momentum transfer efficiency so that more mass may be stored
in disks around low-mass stars.  Best available observational data
suggest $\dot M \propto M_\ast ^2$ (see \S 2.2) but the dependence of
$\Sigma_{\rm d}$ on $M_\ast$ is poorly known. In view of these
uncertainties, we introduce another two series of models.

The parameters of models $x$D and $x$E are identical to those of
models $x$, respectively.  The only difference is that we set 
$h_{\rm d} = M_\ast/M_\odot$ in series D and it is set to be unity for all
$M_\ast$ in series E.  In comparison with the standard models, disks
around low-mass stars are less deficient in these new models while
those around the solar type stars remain the same.

\begin{figure}
\plotone{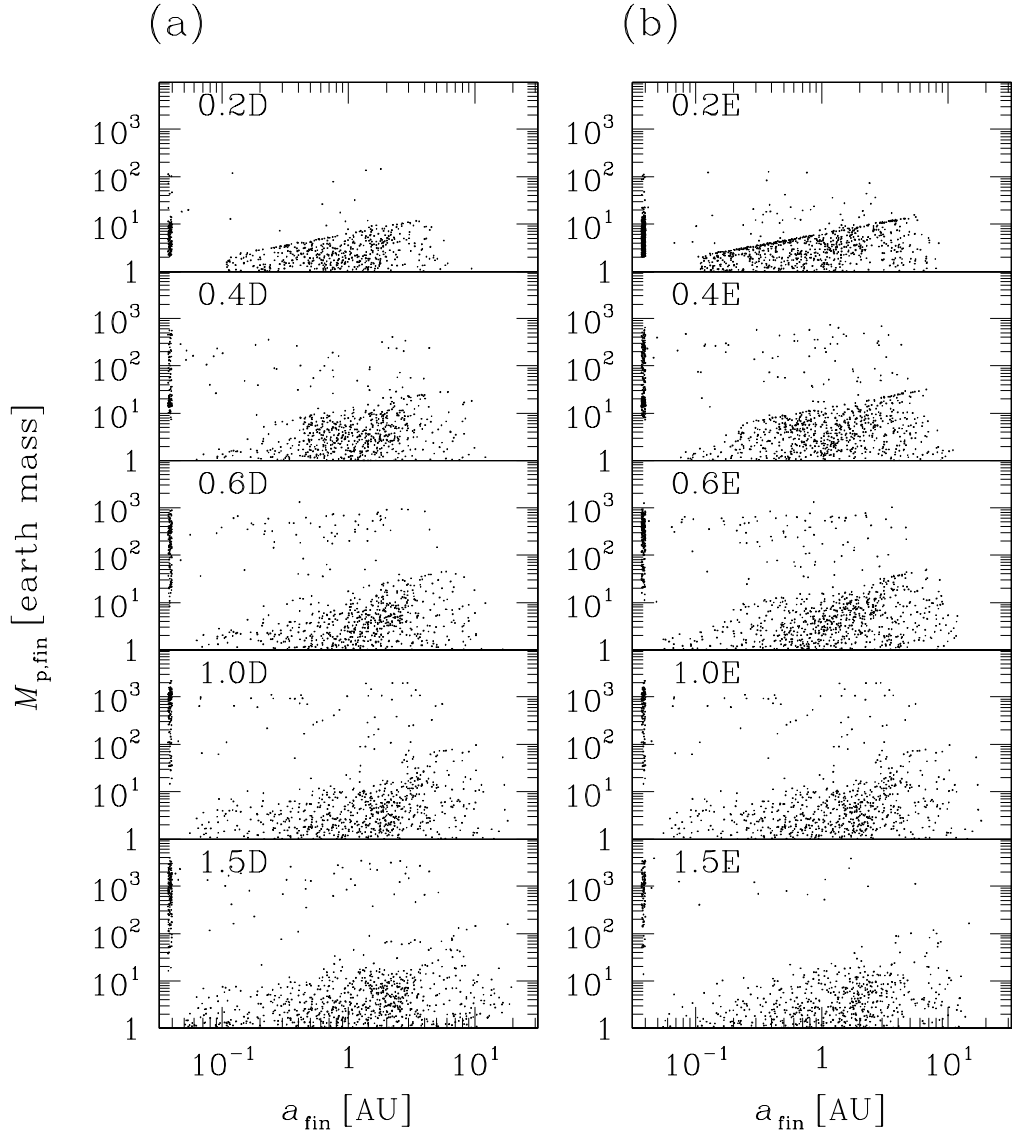}
\caption{
The same as Figure \ref{fig:maBC} except
(a) model $x$D with $h_{\rm d} = M_{\ast}/M_\odot$,
(b) model $x$E with $h_{\rm d} = 1$.
}
\label{fig:maDE}
\end{figure}

The predicted distributions for series D and E are shown
in Figure \ref{fig:maDE}.
Since $M_{\rm p, mig}$ and $M_{\rm p, trunc}$ do not depend on
$\Sigma_{\rm d}$ nor $\Sigma_{\rm g}$, 
planets' migration and gas accretion undergo
the same paths as in standard models for the same mass planets.
As a result, the mass distribution of close-in planets is
self-similar among models $x$, $x$D, and $x$E.
Since massive disks exist around lower mass stars
more frequently in series D (and even more frequently in series E),
the amplitude of mass distributions around lower mass stars
are enhanced in these models.
In series D and E, inferred frequency of Jupiter mass planets
around M stars are comparable to those around G stars
(Figure \ref{fig:maDE}).
The sparse detection of close-in Jupiter-mass and Neptune-mass 
planets around M stars suggests that the disk
masses are rapidly increasing function of $M_\ast$ as we have assumed
in the standard models with $h_{\rm d} = (M_\ast/M_\odot)^2$.  

\section{Summary and discussions}
\label{sec:discussion}

Observational discovery of extrasolar planets is advancing rapidly.
We now have sufficient amount of data to carry out statistical
characterization of planetary properties and to place constraints not
only on the dominant mode of planet formation but also the range of
physical quantities which determine their growth and migration rates.
In this paper, we focused our discussion on the mass function of
close-in planets around stars with various masses, in particular
lower masses than the solar mass, because its origin
is determined by the delicate balance of various processes and they
are the most conspicuous companions of nearby stars.  
Although many ($\ga 90\%$) planets which once migrated to 
the proximity of their host stars may be eliminated (Paper II),
we can compare the mass distribution of close-in planets
among around stars with various masses.
If the elimination factor is taken into account, 
rough comparison is also possible between close-in and more distant
planets. 

Our results are summarized as follows.

\begin{enumerate}

\item Dynamically-isolated, close-in, Neptune-mass planets with silicate
and iron cores can form in relatively massive disks around solar-type
stars.  But their frequency is expected to be an order of magnitude
smaller than that of close-in Jupiter-mass planets.

\item
Dynamically-isolated, close-in,
Neptune-mass ice giants can form in lower-mass stars.  Their
frequency peaks around the M dwarfs.  
Since the luminosity of M dwarfs is weak, the ice boundary is 
located well inside 1 AU.  These planets are formed at
around 1AU but outside the ice boundary and their cores are primarily
composed of volatile ices.  Planets which migrated to the stellar
proximity with masses in the range of $5-15 M_\oplus$ may acquire,
{\it in situ}, a limited amount of additional gas,
but gas accretion is immediately quenched by gap formation
because of small aspect ratio of the disk in the proximity of the host star. 
Because these planets compose mostly of icy material
and M stars' luminosity is relatively weak,
they may have water-vapor atmosphere and water ocean \citep{Leger04}.  

\item
Embryos with mass lower than $10 M_\oplus$ cannot migrate to the
proximity of their F, G, and K dwarf host stars through type II
migration.  Detection of dynamically-isolated Earth-mass close-in
planets may be attributed to type I migration  of
low-mass embryos \citep{Ward97}, sweeping secular resonances
\citep{Lin_etal04}, or sweeping mean motion resonances.
Around late M dwarfs, however, dynamically-isolated, a few earth-mass
planets can form with temperature comparable to that 
of the earth.  

\item
Around M dwarfs, the formation probability of gas giants is
much reduced. The relatively low $\Sigma_{\rm d}$ prevents the
emergence of sufficiently massive cores prior to the severe depletion
of gas in their nascent disks, which is consistent with
the results by \citet{Laughlin04b}. 

\item
The mass function of close-in planets generally have
two peaks at about Neptune mass and at about Jupiter mass.
The lower-mass peak takes the maximum frequency for M stars, while
the higher-mass peak is 
much more pronounced around higher-mass stars (F, G, K dwarfs).
These are because planets tend to undergo type II migration after
fully accreting gas around the higher-mass stars 
while they tend to migrate faster than gas accretion around M stars.
Unless the termination location of
planetary migration is a decreasing function of $M_\ast$, close-in
Neptune-mass planets around M dwarfs are easier to detect than 
those around G dwarfs.

\item
The mass function of dynamically-isolated close-in planets around
stars with various masses can also be used to calibrate the sufficient
condition for the onset of planetary migration and for the termination 
of gas accretion due to planet-disk tidal interaction. 
\end{enumerate}

The metallicity dependence on frequency of extrasolar gas
giant planets may not be easily accounted for by the 
gravitational instability scenario, e.g., \citep{Boss01},
while it is naturally accounted for by the sequential 
core accretion scenario that we are based on (Paper II).
The condition of the gravitational instability is 
$1 > Q = c_s \Omega_{\rm K}/\pi G \Sigma_{\rm g} \propto
M_\ast^{1/2} M_\ast^{1/2}/\Sigma_{\rm g} = M_\ast^{1-\beta}$,
and its radial wavelength
 $\lambda  = 2\pi^2 G \Sigma_{\rm g}/ \Omega_{\rm K}^2
\propto M_\ast^{\beta-1}$ \citep{Toomre64} where
$\Sigma_{\rm g} \propto M_\ast^{\beta}$.
Hence, the instability may be more limited and result in smaller
clumps around lower-mass stars, which 
could account for the above features 4 and 5, if $\beta = 2$ is 
assumed as in the standard series in the present paper.
However, if $\beta = 1$, the gravitational instability scenario
cannot account for the above features 4 and 5,
while the sequential 
core accretion scenario still shows the tendency for the features.
More detailed study on the dependence of disk mass on
stellar mass $M_\ast$ is needed. 

Our work is primarily motivated by the discovery of GJ436b \citep{Butler04}. 
Our theoretical extrapolations can be tested with the following statistical
properties of close-in planets to be discovered by various techniques.

\begin{enumerate}
\item
As mentioned above, the mass function of dynamically-isolated 
close-in planets as a
function of the spectral classes of their host stars is particularly
useful in the determination of the growth, migration, and disk
depletion time scales.

\item
A comparison between the frequencies of gas giants with close-in 
orbits and those with extended orbits can provide constraints on the 
migration condition, survival criteria, and disk mass as functions
of the host stars' mass.

\item
In systems with multiple giant planets, the dynamical architecture may provide 
clues on whether the migration of the close-in planets are driven
by planet-disk tidal interaction or sweeping secular resonance.

\item
A comparison of atmospheric properties of close-in Neptune-mass
planets around G dwarfs to that around M dwarfs can verify the conjecture that
the former have silicate and iron cores whereas the latter have ice
cores.

\end{enumerate}

The following observations of protostellar disks may provide useful input
to the model:

\begin{enumerate}
\item
A spatially resolved image of disks can directly provide information
of $\Sigma_{\rm d}$ and the temperature distribution around any given
host star.

\item
The dependence of $\Sigma_{\rm d}$ on the mass of the host stars
determines the functional form of $h_{\rm d}$.

\item
A relation between the disk mass and accretion rate onto the host stars
places a constraint on the rate of type II migration.

\item
A direct measurement of the gas distribution is particularly important in
determining $\tau_{\rm dep}$ and $h_{\rm g}$.
\end{enumerate}

On the modeling side, we need to consider:

\begin{enumerate}
\item
The possibility of radiative feedback on the termination of gas accretion.

\item
The stoppage of type II migration and the survival of short-period planets.

\item
The rate (and direction) of type I migration of cores.

\item
The enhanced probability of multiple planet formation.

\item
Effect of dynamical interaction between multiple planets during and after
gas depletion.

\item
The radial distributions of $\Sigma_{\rm d}$ and $\Sigma_{\rm g}$.
Effects of different power-law index of $a$ dependence from
$-1.5$ that we used. More realistic time evolution of $\Sigma_{\rm g}$.

\item
The influence of a stellar companion on the emergence and survival of
planets.

\end{enumerate}

Some of these issues will be addressed in future discussions.  

\acknowledgements We thank G. Laughlin, M. Nagasawa, G. Ogilvie, and
S. Vogt for useful discussions, and the anoymous referee for
useful comments.  This work is supported by the NASA
through NAGS5-11779 under its Origins program, JPL 1228184 under its
SIM program, and NSF through AST-9987417 and by JSPS.

\clearpage

\clearpage

\clearpage

\clearpage

\clearpage

\clearpage

\clearpage

\clearpage

\clearpage

\end{document}